%% file: 00-all.tex
\documentclass{article}
\usepackage{graphicx} 
\usepackage{amsmath, amsthm, amssymb}
\usepackage{mathtools}
\usepackage{indentfirst}
\usepackage[left=1in, bottom=1in, right=1in, top=1in]{geometry}
\usepackage[authoryear, square]{natbib}
\usepackage{xcolor}
\usepackage{algorithm}
\usepackage{algorithmic}
\usepackage[toc,page]{appendix}
\usepackage{subcaption}
\usepackage{graphicx}
\usepackage{authblk}
\usepackage{footmisc}

\DeclareMathOperator*{\argmin}{arg\,min}
\numberwithin{equation}{section}
\newtheorem{proposition}{Proposition}[section]

\title{A Pathwise Coordinate Descent Algorithm for \\ LASSO Penalized Quantile Regression}
\author[1]{Sanghee Kim \thanks{Email: sk2689@cornell.edu}}
\author[1]{Sumanta Basu \thanks{Email: sumbose@cornell.edu}}

\affil[1]{Department of Statistics and Data Science, Cornell University}

\begin{document}

\maketitle

\begin{abstract}
$\ell_1$ penalized quantile regression is used in many fields as an alternative to penalized least squares regressions for high-dimensional data analysis. Existing algorithms for penalized quantile regression either use linear programming, which does not scale well in high dimension, or an approximate coordinate descent (CD) which does not solve for exact coordinatewise minimum of the nonsmooth loss function. 
Further, neither approaches build fast, pathwise algorithms commonly used in high-dimensional statistics to leverage sparsity structure of the problem in large-scale data sets. To avoid the computational challenges associated with the nonsmooth quantile loss, some recent works have even advocated using  smooth approximations to the exact problem. In this work, we develop a fast, pathwise coordinate descent algorithm to compute exact $\ell_1$ penalized quantile regression estimates for high-dimensional data. We derive an easy-to-compute exact solution for the coordinatewise nonsmooth loss minimization, which, to the best of our knowledge, has not been reported in the literature. We also employ a random perturbation strategy to help the algorithm avoid getting stuck along the regularization path. In simulated data sets, we show that our algorithm runs  substantially faster than existing alternatives based on approximate CD and linear program, while retaining the same level of estimation accuracy.
\end{abstract}

\input{01-intro}

\input{02-background}
\input{03-method}
\input{04-simulation}
\input{05-conclusion}

\newpage
\appendix
\input{06-appendix}

\end{document}

%% file: 01-intro.tex
\section{Introduction}
\label{sec:intro}
\citet{koenker1978regression} first introduced quantile regression (QR) estimator as a robust  alternative to least squares estimator. Beyond  robustness, QR estimators can describe the entire conditional distribution of the response variable given the predictor variables. This property can be used to understand the heterogeneous effects of predictors on response by studying how the conditional distribution changes across quantiles  \citep{hao2007quantile}. In high dimension, one can select important predictor variables that have large effects on the response variable or account for the heteroskedasticity of the large dataset by using penalized quantile regression (PQR). For example, \citet{nascimento2017regularized} applied PQR to select genome to improve the prediction of genomic estimated breeding values (GEBV). \citet{bayer2018combining} uses PQR to combine the predicted Value at Risk (VaR) when multicollinearity is present. \citet{nguyen2020investigating} obtains the relevant drivers of the VaR of a cryptocurrency through PQR and construct a VaR network. As the dimension of the problem grows, computation of the QR estimator over a large grid of penalty parameters becomes quickly prohibitive, and there is a need for scalable algorithms. \\

We consider the problem of computing $\ell_1$-penalized quantile regression ($\ell_1$-QR) estimators for high-dimensional data sets. Formally, given data points $(x_1, y_1), \ldots, (x_n, y_n)$, where $y_i \in \mathbb{R}$ is a numerical response variable, and $x_i = (x_{i1}, x_{i2}, \ldots, x_{ip})^\top \in \mathbb{R}^p$ is a $p$-dimensional covariate, we solve the following optimization problem

\begin{equation}\label{eqn:qr-lasso}
\argmin_{\beta \in \mathbb{R}^p} \sum_{i=1}^n \rho_\tau \left(y_i - x_i^\top \beta \right) + \lambda \sum_{j=1}^p \left| \beta_j \right|,
\end{equation}
where $\rho_\tau \left(u\right):= u \left(\tau-\mathbf{I}(u<0)\right)$ is the \textit{check loss} function with indicator function $\mathbf{I}(.)$, and $\lambda$ is a penalty parameter to be chosen in a data-driven fashion. It is common to solve the above problem on a dense grid of values of $\lambda$, and use some form of model selection criterion (such as AIC, BIC) or cross-validation (CV) to select the optimal $\lambda$. In high-dimensional statistics with squared error loss, \textit{pathwise coordinate descent} algorithms are state-of-the-art in terms of speed and accuracy for computing the solutions over a grid of $\lambda$. In \citet{friedman2007pathwise}, the computational cost of coordinate descent was shown to be much lower than least angle regression (LARS). In this work, we develop a similar algorithm for \textit{check loss} functions instead of squared error loss. \\

Traditionally, $\ell_1$-QR estimators are computed using linear programming (LP) techniques. \citet{koenker2005quantile} suggested interior-point method which searches for the optimal solution within a polytope that is defined using a system of linear equations. \citet{li20081} also formulates $\ell_1$-QR as a linear program and gives the entire solution path. Another algorithm utilized by \citet{gu2018admm} is the alternating direction method of multipliers (ADMM) where the main problem is broken into sub problems by using the Lagrangian to obtain the global solution. \citet{peng2015iterative} proposed iterative coordinate descent (QICD) and reduced the main problem into a sequence of weighted median finding problem, similar to that of the $\ell_1$-penalized least absolute deviations (LAD) in \citet{wu2008coordinate}. All the above methods are known to be efficient, but share a common weakness when computing the solution over a grid of penalty $\lambda$'s. That is, they do not use the sparsity structure of the problem along the penalty grid to reduce the computational burden. \\

In high-dimensional statistics, similar issue with $\ell_1$-penalized regression problems with squared error loss are overcome with $\textit{pathwise}$ coordinate descent (CD) algorithms due to their speed and accuracy. This is because of two main reasons. First, the individual coordinatewise minima have closed form updates (soft thresholded inner product between predictor and partial residuals) and are very easy to compute. Second, these algorithms can leverage problem-specific sparsity structure through strategies such as warm start, active set selection, and screening rules - a feature that most off-the-shelf convex optimization algorithms do not share. These pathwise CD also come with some convergence guarantees. According to \citet{tseng2001convergence}, the CD solution of a convex, separable $\ell_1$-penalized least squares problem would converge to a stationary point under some regularity conditions. \\

There are several challenges to directly adopt the same strategy in our case. Since the check loss function is nonsmooth and non-differentiable, there is no closed form update for coordinatewise minimum in the setup of \eqref{eqn:qr-lasso} known in the literature. Instead, recent works have changed the original check loss \eqref{eqn:qr-lasso} to a different computationally tractable loss function. For example, \citet{yi2017semismooth} combined CD with semismooth Newton method to solve the nonsmoothness. Their method approximates the quantile loss with Huber loss and employs semismooth Newton coordinate descent (SNCD). In order to overcome non-differentiability, \citet{fernandes2021smoothing} developed convolution smoothed quantile regression (SQR) method by applying kernel function to get a convex, twice differentiable loss function. Then, one could solve SQR with CD and ADMM presented by \citet{tan2022high}. The absence of a closed form solution also complicates developing fast screening rules that leverage underlying sparsity structure of the problem. Moreover, these smooth approximation are often sensitive to additional tuning parameters that govern the amount of smoothing.\\

\begin{figure}[t!]
\begin{center}
\centerline{\includegraphics[width=\columnwidth]{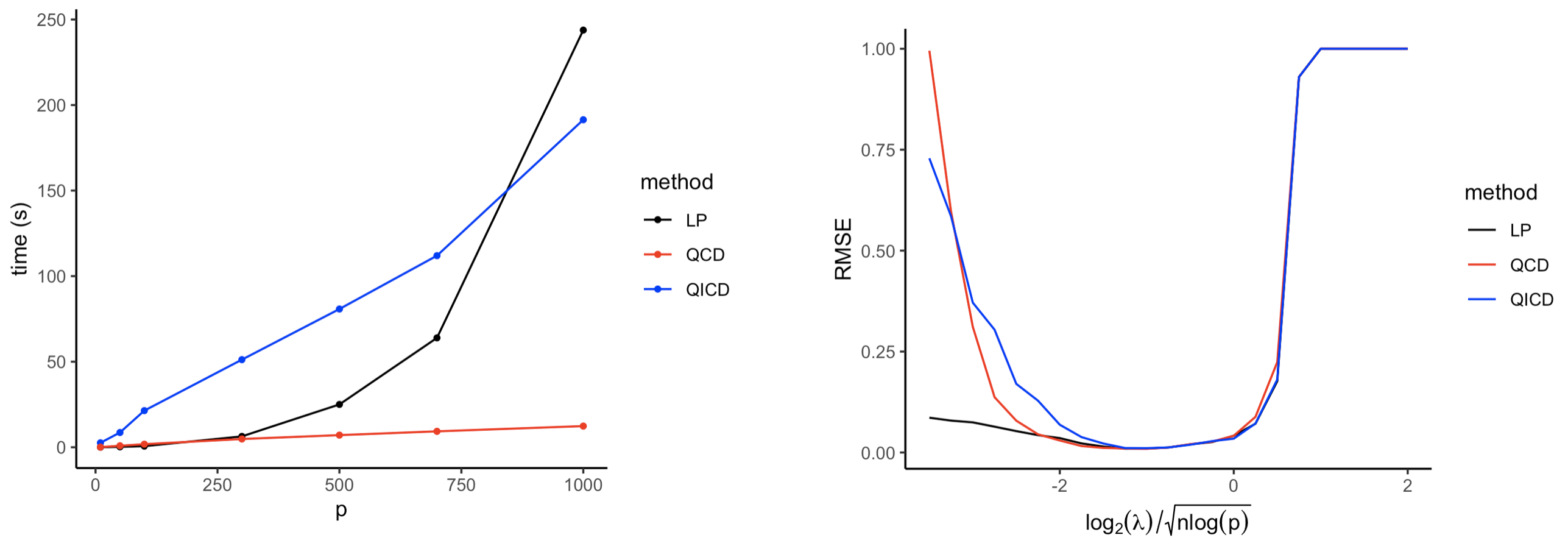}}\label{fig:exa}
\caption{\textit{An example of runtime for different dimensions and regularization path in high dimension.} QCD (red line) is the fastest and achieves the same amount of accuracy compared to LP and QICD (details in Section \ref{sec:simulation}). (left) Runtime of LP, QICD, and QCD for dimensions $p=10, 50, 100, 300, 500, 700, 1000$ with fixed $n=300$. (right) Regularization path of the three methods when $p=1000$ and $n=300$.} 
\label{fig:motivation}
\end{center}
\vskip -0.2in
\end{figure}

In this work, we make several contributions to overcome the above challenges and develop a fast, pathwise CD (QCD) algorithm to solve the original problem \eqref{eqn:qr-lasso}. First, we derive an easy-to-compute solution for coordinatewise minima which can be viewed as finding a quantile of $n$ scalars. Our solution leverages the fact that the partial derivative of \eqref{eqn:qr-lasso} is piecewise constant and monotone increasing. We track the derivative over the real line and search for zero-crossing. This operation has the same complexity as finding a weighted median of $n$ real numbers. Second, we observe that checking whether the coordinatewise minima is zero or non-zero is much cheaper than finding the exact minima, and is equivalent to calculating a single inner product. This opens up the possibility of explicitly leveraging the underlying sparsity structure of the problem and dramatically reduces the computational burden in high-dimensional sparse problems as shown in Figure \ref{fig:motivation}. Third, to avoid the algorithm from getting stuck along the regularization path, we introduce a random perturbation, \textit{nudge}. We add a random noise to the estimated $\hat{\beta} \in \mathbb{R}^p$ before proceeding to the next $\lambda$ along the regularization path. Finally, we provide a fast implementation of our algorithm in an R package with its inner layer coded in FORTRAN 77. \\

Using various simulated data, we benchmark our algorithm QCD against two methods: LP which uses interior-point method (\texttt{quantreg} package in R, \citet{koenker2005quantile}) and QICD which gives approximate coordinate descent update (\texttt{QICD} package in R, \citet{peng2015iterative}). We compare the runtime and relative mean squared error (RMSE) of these algorithms along the regularization path. A small example of the runtime and RMSE plot is shown in Figure \ref{fig:motivation}. QCD is the fastest in terms of computational time as the dimension of the problem increases and its RMSE is as low as the RMSE of QICD or LP in high dimension. \\

The rest of the paper is organized as follows. Section 2 gives a brief review of the QR and PQR literature. In Section 3, we present our exact CD update for $\ell_1$-QR. In Section 4, we benchmark the performance of our algorithm on simulated data sets. In the Appendix, we show how our exact CD update can be generalized to nonconvex smoothly clipped absolute deviation (SCAD) and minimax concave penalty (MCP) penalties. Our R package \texttt{QCD}, which uses a FORTRAN 77 implementation of our algorithm, is available at \texttt{https://github.com/sangheekim96/QCD}.

%% file: 02-background.tex
\section{Background}\label{sec:background}

In this section, we provide a summary of computational aspects of penalized quantile regression. Before we begin, we provide a short description of calculating quantiles in classical statistics, and illustrate the strategy of derivative tracking that is at the heart of our algorithm. \\

\begin{figure}[t!]
\begin{center}
\centerline{\includegraphics[width=\columnwidth]{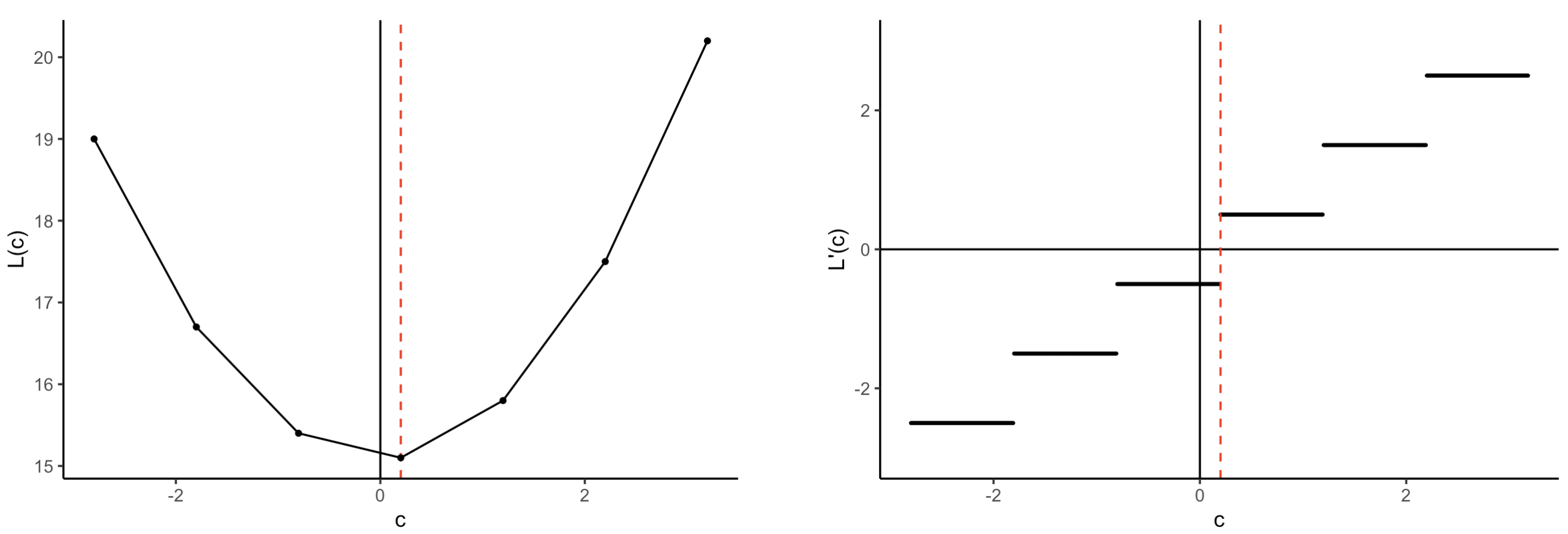}}
\caption{\textit{Objective function $L(c)$ and its derivative $L'(c)$ for check loss minimization of \eqref{eqn:qr} with $\tau=0.5$.} The minimum of $L(c)$ is achieved at the median, the red dashed line at $c=0.2$, which is exactly where the derivative first crosses zero.}
\label{fig:simpleQR}
\end{center}
\vskip -0.2in
\end{figure}

The $\tau$th sample quantile 
is the solution to the problem 
\begin{equation}\label{eqn:qr}
     \argmin_{c \in \mathbb{R}} L(c) = \argmin_{c \in \mathbb{R}} \sum_{i=1}^n \rho_\tau (y_i - c),
\end{equation}
where $\rho_\tau \left(u\right):= u \left(\tau-\mathbf{I}(u<0)\right)$ is the so-called \textit{check loss} function. Note that the derivative $L'(c)$ exists everywhere except at $y_1, \ldots, y_n$, and the solution of \eqref{eqn:qr} can be obtained by tracking when the derivative 
\begin{equation}\label{eqn:qr_Lprime}
    L'(c) = \sum_{i \in \{ i : y_i \geq c \} } (-\tau) + \sum_{i \in \{ i : y_i < c \} } (1-\tau) 
\end{equation}
crosses $0$. In Figure \ref{fig:simpleQR}, we display the plots of $L(c)$ and $L'(c)$ to demonstrate that finding median $(\tau=0.5)$ is exactly equivalent to finding when $L'(c)$ crosses zero. This observation motivates our strategy to solve the non-smooth optimization problem with piece-wise constant derivatives.\\

Now consider a higher dimension general QR problem with $p$ variables 
\begin{equation} \label{eqn:qr-pdim}
    \min_{\boldsymbol{\beta} \in \mathbb{R}^p} \ \sum_{i=1}^n \rho_\tau (y_i - x_i^\top \beta). 
\end{equation}
We can form a penalized QR problem 
\begin{equation} \label{eqn:pqr}
    \min_{\boldsymbol{\beta} \in \mathbb{R}^p} \ \sum_{i=1}^n \rho_\tau (y_i - x_i^\top \beta) + \sum_{j=1}^p p_\lambda(|\beta_j|) 
\end{equation}
by adding a penalty function $p_\lambda(|\beta_j|)$ to perform variable selection that yields a sparse solution. The choice of penalty function varies from convex $\ell_1$ (LASSO) penalty \cite{tibshirani1996regression}, where $p_\lambda(|\beta|) = \lambda |\beta|$, to non-convex smoothly clipped absolute deviation (SCAD) and minimax concave penalty (MCP) \cite{fan2001variable, zhang2010nearly}.\\

Traditionally, QR estimators are computed using LP solvers: simplex \cite{barrodale1974algorithm} and interior-point methods \cite{koenker1996quantile} by framing \eqref{eqn:qr-pdim} as a linear program. In presence of $\ell_1$ penalty, the problem can still be solved with these methods since the structure of linear program is maintained. However, the computational cost of LP becomes expensive in large scale datasets where $p$ becomes as large as $n$ or larger than $n$ \cite{he2023smoothed}. \\

Coordinate descent (CD) is a competitive alternative to LP for solving both low- and high-dimension penalized problems due to its flexibility of handling both convex and non-convex penalty \cite{edgeworth1887observations, li2004maximum, friedman2007pathwise,  wu2008coordinate, peng2015iterative}. In particular, the QICD algorithm \cite{peng2015iterative} solves non-convex penalized QR problem through iterative CD. We briefly describe the QICD algorithm to solve $\ell_1$ penalized quantile regression problem ($\ell_1$-QR) by applying the same logic described in the paper and refer to \citet{peng2015iterative} for details. When iterating until convergence to obtain a set of vector $\left( \hat{\beta_1}, \ldots, \hat{\beta_p} \right)$, the estimated $\hat{\beta_j}^{(r+1)}$, $j = 1, \ldots, p$, at iteration $r+1$ can be written as 
\begin{align} \label{eqn:qicd}
    \hat{\beta_j}^{(r+1)} &:= \argmin_{\beta_j} \left\{ \frac{1}{n+1} \sum_{i=1}^{n+1} w_{ij} |u_{ij}| \right\}, 
\end{align}
where $u_{ij}$ and $w_{ij}$ are defined as  follows:
\begin{align*}
    u_{ij} &:= \begin{dcases}
			\frac{y_i - \sum_{s<j} x_{is} \hat{\beta_s}^{(r+1)} - \sum_{s>j} x_{is} \hat{\beta_s}^{(r)}}{x_{ij}} - \beta_j, & i=1, \ldots, n \\
            \beta_j, & i=n+1,
		 \end{dcases} \\
         w_{ij} &:= \begin{dcases}
    \frac{1}{n} \left| x_{ij} \left( \tau - \mathbf{I}(x_{ij} u_{ij} < 0) \right) \right|, & i=1, \ldots, n \\
    \lambda, &i=n+1.
    \end{dcases}
\end{align*}

It is important to note that when solving for this coordinatewise minimum, QICD alters the actual objective function and yield inexact CD solution because the weights $w_{ij}$ rely on $\beta_j$ through $u_{ij}$, and $\hat{\beta}_j^{(r)}$ is plugged in here so that \eqref{eqn:qicd} can be solved by finding a weighted median. We provide a short illustration of the impact of this approximation in Figure \ref{fig:converge}, with complete description in Appendix \ref{apdx:qcd_conv}. As a result, this approximation complicates the convergence analysis of the algorithm from \citet{tseng2001convergence}, which relies on finding \textit{exact} coordinatewise minimum in each CD iteration. \\
\begin{figure}[t!]
\begin{center}
\centerline{\includegraphics[width=\columnwidth]{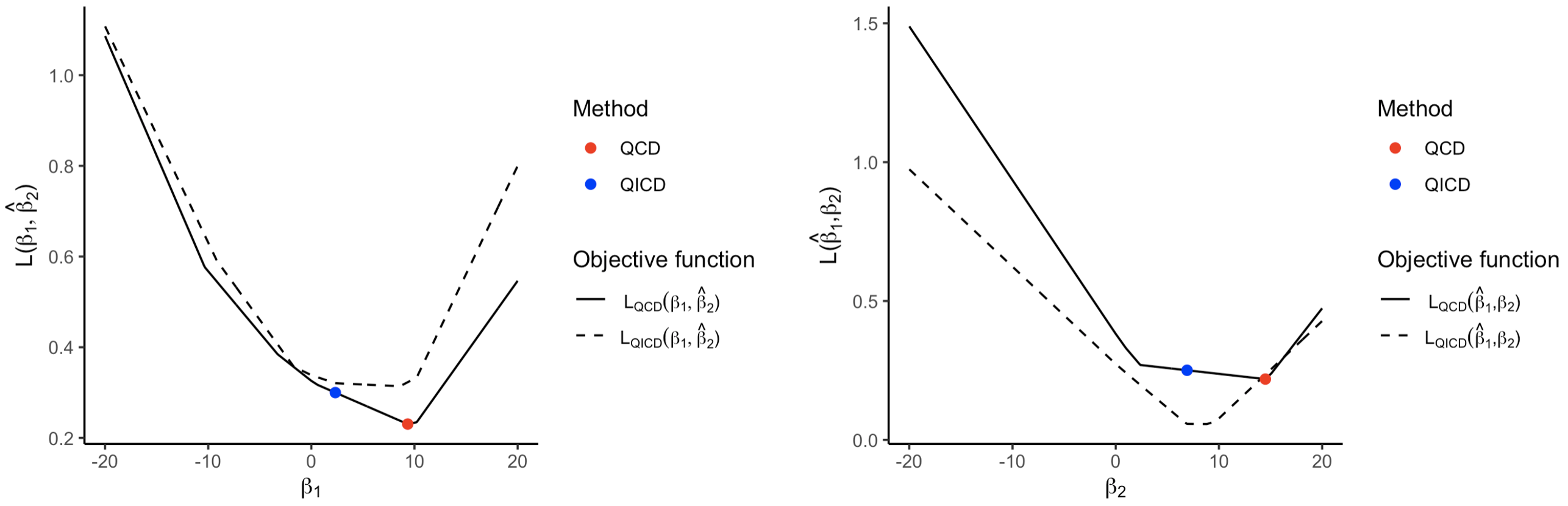}}
\caption{\textit{An example (details in Appendix \ref{apdx:qcd_conv}) where QICD updates fail to achieve coordinatewise minimum.} In two iterations, the objective function of QCD (solid line) is different from the objective function of QICD (dashed line) and achieves minimum in different places.}
\label{fig:converge}
\end{center}
\vskip -0.2in
\end{figure}

In contrast to QICD that focuses on a single  $\lambda$ in $\ell_1$-QR, we consider a grid of parameters $\lambda \in \left[ \lambda_{min}, \lambda_{max} \right]$ and perform \textit{pathwise} CD. This allows us to explicitly study the behavior on the \textit{entire} regularization path. Our pathwise algorithm currently uses \textit{warm start}, where we start from $\lambda_{\max}$ and decrease the $\lambda$ value gradually while initializing the algorithm at every step with the previous solution along the grid. Our algorithm is also amenable to employ active set selection where only a fraction of coefficients are estimated throughout the iteration \cite{friedman2007pathwise, friedman2010regularization, hastie2015statistical}, although it is not currently implemented in our software. We refer to Appendix \ref{apdx:background} for more detailed background description of methods to solve quantile regression problems.

%% file: 03-method.tex
\section{Method}\label{sec:method}
\begin{figure}[t!]
\begin{center}
\centerline{\includegraphics[width=
0.5\columnwidth]{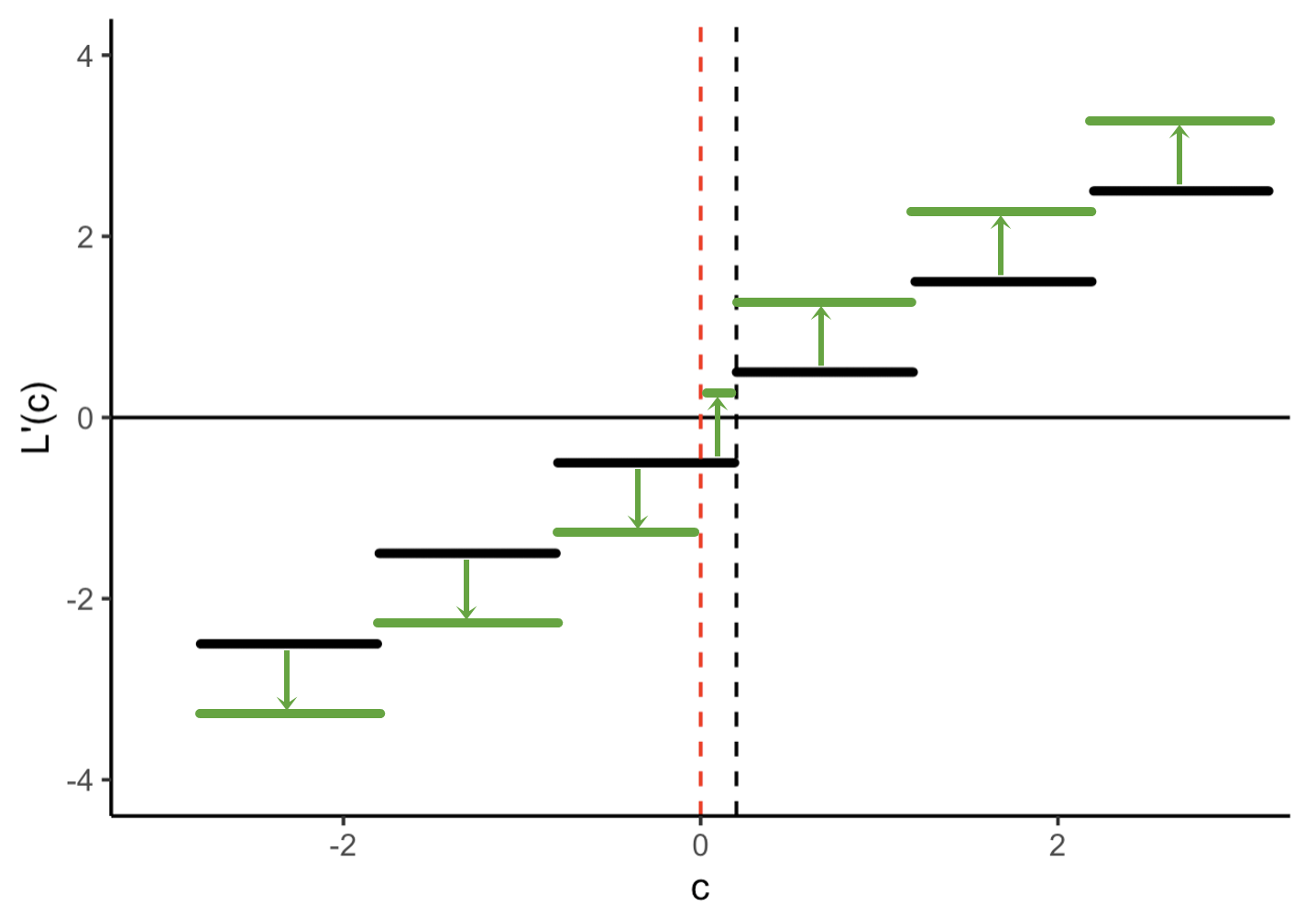}}
\caption{\textit{Simulation example of the derivative $L'(c)$ for minimization of \eqref{eqn:qr_l1} with $\tau=0.5$.} The black dashed line represents the original median $c=0.2$ and the red dashed line represents the new median $c=0$ after the boost.}
\label{fig:simpleQR_l1}
\end{center}
\vskip -0.2in
\end{figure}

In this section, we demonstrate our derivative tracking algorithm, QCD, by deriving an exact CD update for $\ell_1$-QR problem. Before describing QCD algorithm, we motivate $\ell_1$-QR by recalling the simplest QR problem \eqref{eqn:qr} 
\begin{align}
\label{eqn:qr_l1}
    \argmin_{c \in \mathbb{R}} L(c) = \argmin_{c \in \mathbb{R}} \sum_{i=1}^n \rho_\tau (x_i - c) + \lambda |c|,
\end{align}
and the corresponding derivative $L'(c)$ as
\begin{align}\label{eqn:qr_l1_Lprime}
    L'(c) = \sum_{i \in \{ i : x_i \geq c \} }(-\tau) + \sum_{i \in \{ i : x_i < c \} }(1-\tau) + \lambda \times sign(c)
\end{align}
where $sign(c) \in \{-1,0,1\}$ depends on the sign of $c$. Since $\lambda \times sign(c)$ is also piecewise linear, we only need to add $+\lambda$ or $-\lambda$ to \eqref{eqn:qr_Lprime} in order to obtain $L'(c)$ in \eqref{eqn:qr_l1_Lprime}. The new derivatives are depicted in Figure \ref{fig:simpleQR_l1} with green lines. We give $-\lambda$ boost to the derivatives when $c<0$ and give $+\lambda$ boost when $c>0$. We see that the median is shifted from $0.2$ to $0$ since the boost encourages $L'(c)$ to cross zero at exactly $c = 0$ instead of $c=0.2$. This operation of penalized median finding can be viewed as a form of soft-thresholding that shrinks the optima to $0$ when the true optima is small compared to the strength of the penalty $\lambda$. Building upon this intuition, we next  describe step-by-step how to compute the coordinatewise minima for our QCD algorithm. \\

\textit{\textbf{Computing exact coordinatewise minimum}}. Let us define the scaled partial residual for the $j^{th}$ predictor as $v_{ij}:= (y_i - \sum_{k \neq j} x_{ik} \hat{\beta}_k)/x_{ij}$, for $i=1, \ldots, n$, and $j = 1, \ldots, p$. Then  we can formulate the $j^{th}$ coordinatewise minimization problem in \eqref{eqn:qr-lasso} as 
\begin{align} \label{obj_qrlasso}
    &\min_{\beta_j} \ \sum_{i=1}^n \rho_\tau \left( x_{ij} (\underbrace{\frac{y_i - \sum_{k \neq j} x_{ik} \hat{\beta}_k}{x_{ij}}}_{v_{ij}} - \beta_j) \right) +  \lambda |\beta_j| \nonumber \\
    \Leftrightarrow &\min_{\beta_j} \sum_{i=1}^n \rho_\tau \left(x_{ij}(v_{ij} - \beta_j) \right) + \lambda|\beta_j| \nonumber \\
    \Leftrightarrow &\min_{\beta_j} \sum_{i : x_{ij}>0} |x_{ij}|\rho_\tau(v_{ij}-\beta_j) 
 + \nonumber \\ & \ \ \ \ \ \ \ \sum_{i : x_{ij}<0} |x_{ij}|\rho_{1-\tau}(v_{ij} - \beta_j) + \lambda|\beta_j|
\end{align}
for each $j=1, \ldots, p$. Here, we use a few properties of check loss function, $\rho_\tau \left(u\right):= u \left(\tau-\mathbf{I}(u<0)\right)$. That is, $\forall k>0$, $\rho_\tau(ku)=k \rho_\tau(u)$, $\rho_\tau(-ku)=k \rho_\tau(-u)$, and $\rho_\tau(-u)=\rho_{1-\tau}(u)$. \\

For ease of exposition, in the remaining part of this section we assume $j$ as fixed, and with a little abuse of notation drop the suffix $j$ to write $x_{i}$ instead of $x_{ij}$, $v_i$ instead of $v_{ij}$, $\beta$ instead of $\beta_j$, and simplify \eqref{obj_qrlasso} for one $\beta$:
\begin{equation} \label{eqn:obj_qrlasso_simple}
      \min_{\beta} \sum_{i : x_i>0} |x_{i}|\rho_\tau(v_{i}-\beta)+ \sum_{i : x_i<0} |x_{i}|\rho_{1-\tau}(v_{i}-\beta) + \lambda|\beta|
\end{equation}
Note that \eqref{eqn:obj_qrlasso_simple} is non-differentiable at $v_1, \ldots, v_n$,  $v_{n+1}:= 0$, and constant everywhere else.
Next step is to sort $n+1$ observations $v_1, \ldots, v_n, v_{n+1}$ and compute the derivative of the objective function in \eqref{eqn:obj_qrlasso_simple} recursively. Without loss of generality, assume the sorted sequence to be $\{ v_1, v_2, \ldots, 0, \ldots, v_{n+1} \}$. After sorting, we sequentially consider $\beta$ within each segment of the form $(v_j, v_{j+1})$ and compute the derivative. As in Figure \ref{fig:simpleQR_l1}, the gradient of $\beta$ within each segment is constant, and can be calculated recursively, as shown in the proposition below. 
\begin{proposition}\label{prop1}
Define $S' = -\sum_{i=1}^{n} |x_i|  \left( \tau \mathbf{I}(x_i > 0) + (1-\tau)\mathbf{I}(x_i < 0)  \right)$. Then, the derivative when $v_m < \beta < v_{m+1} < \ldots < 0$ can be expressed as 
$$S_m' = S' + \sum_{i=1}^m |x_i| - \lambda.$$
The derivative when $\beta$ exceeds $0$, $0 < \ldots <v_q < \beta < v_{q+1}$, can be expressed as
$$S_{q}' = S' + \sum_{i=1}^q |x_i| + \lambda.$$
\end{proposition}

Detailed proof is given in Appendix \ref{apdx:derivative_proof}. Applying Proposition \ref{prop1}, we can obtain all the derivatives and define the exact coordinatewise minimum in the next proposition.
\begin{proposition}\label{prop2}
    The solution of \eqref{eqn:obj_qrlasso_simple} is given by 
    $$\hat{\beta} = v_{t},$$ 
    where $t$ is the index satisfying $S_{t-1}' <0$ and $S_t' \ge 0$.
\end{proposition}

The proof follows by combining the following facts. The derivatives are monotone increasing and piecewise constant, and the objective function is continuous. So the function decreases linearly until the derivative crosses zero, then increases linearly. Hence the point where the derivative crosses zero is a minimum.  In other words, the solution $\hat{\beta}$ is equivalent to the scaled partial residual (or $0$) where the derivative crossed zero. 

To estimate $p$-dimensional vector $\beta = (\beta_1, \ldots, \beta_p)$, we need to repeat the solution finding in Proposition \ref{prop2} for each coefficient $\beta_j$. We present the entire steps of QCD in Algorithm \ref{alg1}. Note that we also use Gauss-Seidel scheme, where we use the most recent estimates for each $\beta_{j-1}$, as written in \eqref{eqn:qicd} and \citet{peng2015iterative}. \\

\textit{\textbf{Checking for exact zero coordinatewise minimum}}. In order to increase computational speed, we not only implement warm start but also an additional step that checks whether the Karush-Kuhn-Tucker (KKT) condition is met for the optima to be exactly zero.

\begin{proposition}\label{prop3}
The minimizer of \eqref{eqn:obj_qrlasso_simple} is zero if and only if  
    $$\left|S' + \sum_{i : v_{i}<0} |x_{i}| \right| \leq \lambda.$$
\end{proposition}
If the condition is met, there is no need to sort the partial residuals in Algorithm \ref{alg1}. The proof is given in Appendix \ref{apdx:derivative_proof}. \\  

\textit{\textbf{Random nudge along regularization path}}. This pathwise CD algorithm is fast, but our numerical experiments (details in Section \ref{sec:simulation}) suggest that if we use the same initializer for all $\lambda$, the solutions for very similar $\lambda$'s may be quite different. Using a warm start, i.e. using $\hat{\beta}^{(\ell-1)}$ to initialize the solution $\hat{\beta}^{(\ell)}$, resolves this issue.  However, the algorithm tends to get stuck along the regularization path, especially for small values of $\lambda$, and produces the same solution for different $\lambda$ values. In Figure \ref{fig:qcd_qicd_high}, we show that this behavior is not specific to our QCD algorithm and is observed for QICD as well. \\

To resolve this issue, we add a \textit{nudge} to our QCD algorithm. Nudge is a small random noise that helps pathwise CD get unstuck along the regularization path. We add a nudge to the estimated $\hat{\beta}^{(\ell-1)}$ and use it as a warm start for $\hat{\beta}^{(\ell)}$. In our empirical analysis, a random $N(0, 0.01)$ noise, produces robust result across all simulation scenarios.\\

\textit{\textbf{Convergence of QCD algorithm}}. A complete convergence analysis of QCD is challenging because the check loss function is nonsmooth as well as nonseparable in its coordinates, limiting the applicability of existing convergence results for general coordinate descent algorithms. We do not investigate this issue here, and instead refer the readers to the comprehensive convergence analysis of QICD in \citet{peng2015iterative}. Their analysis relies on the general results for block coordinate descent in \citet{tseng2001convergence}, which ensures convergence to a stationary point assuming the objective function is regular. Our algorithm deviates from QICD in two main directions. First, we compute \textit{exact} coordinatewise minimum, which allows us to apply the results of \citet{tseng2001convergence}. Second, we focus only on convex functions and LASSO penalty as opposed to the nonconvex optimization problems analyzed in \citet{peng2015iterative}, which allows us to carry out analysis under less stringent assumption. For example, assuming the objective function is regular and there is at most one coordinatewise minimum, we can directly invoke Theorem 4.1(c) of \citet{tseng2001convergence} to ensure convergence to a stationary point. The assumption of at most one coordinatewise minimum in that theorem will be satisfied when the predictors $x_i$ are random variables with a continuous distribution, since the derivative in each segment (see Proposition \ref{prop1}) will be non-zero almost surely. However, since the check loss function is not differentiable everywhere, a thorough analysis of the regularity assumption in the above theorem will need more work and beyond the scope of this work.

\begin{algorithm}[t]
   \caption{QCD algorithm for $\ell_1$-penalized QR}
   \label{alg1}
\begin{algorithmic}
   \STATE {\bfseries Input:} $X = \left(x_1 : x_2 : \ldots : x_n \right)^\top \in \mathbb{R}^{n \times p}$ \\ \smallskip \ \ \ \ \ \ \ \ \ \ \ \ \ $y = (y_1, \ldots, y_n) \in \mathbb{R}^n$ \\ \smallskip \ \ \ \ \ \ \ \ \ \ \ \ \ $\tau \in \left(0,1\right)$ \\ \smallskip \ \ \ \ \ \ \ \ \ \ \ \ \ $\lambda = \left(\lambda_1, \ldots, \lambda_L \right) \in \mathbb{R}^L$
   \STATE {\bfseries Result:} $\hat{\beta} = \left( \hat{\beta}^{(1)} : \hat{\beta}^{(2)} : \ldots : \hat{\beta}^{(L)} \right)^\top \in \mathbb{R}^{L \times p}$
   \STATE Initialize $\beta \gets \beta^{(0)} = \left( 0, \ldots, 0 \right)^\top$

   \FOR{$\ell=1$ {\bfseries to} $L$}
    \FOR{$j=1$ {\bfseries to} $p$}
        \STATE Compute 
        \[ v_{ij} = \begin{dcases*} 
    \frac{y_i-{\sum_{k \neq j} x_{ik}\hat{\beta}_k}}{x_{ij}}  & $i=1, \ldots, n$ \\ 0 & $i=n+1$ \end{dcases*}\]

    \STATE Define $S'$ as Proposition \ref{prop1}
    
    \IF{$\left|S' + \sum_{i : v_{ij}<0} |x_{ij}|\right| \leq \lambda_\ell$}
        \STATE $\beta_j \gets 0$ \\
      $j \gets j+1$
      
    \ELSE{
    \STATE Order $\{x_{ij}\}_{i=1}^n$ in increasing order of $\{v_{ij}\}_{i=1}^n$  
    
    \STATE Sort $\{v_{ij}\}_{i=1}^n$ in increasing order
    }
   \ENDIF

    \STATE Compute derivatives $S_1', \ldots, S_q'$ in Proposition \ref{prop1} 

    \STATE $\beta_j \gets \hat{\beta}_j=v_{t,j}$ as Proposition \ref{prop2} \\ $j \gets j+1$
   \ENDFOR


    \IF{nudge = TRUE}
    \STATE $\beta \gets \hat{\beta}^{(\ell)} + $ nudge
    \ENDIF

    \STATE $\beta \gets \hat{\beta}^{(\ell)}$
   \ENDFOR
\end{algorithmic}
\end{algorithm}

%% file: 04-simulation.tex
\section{Numerical Experiments}\label{sec:simulation}

In this section, we conduct simulation studies to evaluate the performance of QCD (Algorithm \ref{alg1}), benchmarking with existing methods. 
We follow the simulation setting of \citet{peng2015iterative} to compare QICD with our QCD. 
We first generate $\left( \Tilde{X}_1, \Tilde{X}_2, \ldots, \Tilde{X}_p \right)^\top$ from multivariate normal distribution $N_p (0, \Sigma)$ where $\Sigma = \left((\sigma_{ik})\right)_{p \times p}$ is a covariance matrix with $\sigma_{ik}=0.5^{|i-k|}$. Then, define predictor variables $X_1 = \Phi(\Tilde{X}_1)$ and $X_i = \Tilde{X_i}, \ i=2,3,\ldots, p$. Here, $\Phi(\cdot)$ is the cumulative distribution function of standard normal distribution. The model for generating the response variable is
\begin{equation*}
    Y = X_6 + X_{12} + X_{15} + X_{20} + 0.7 X_1 \epsilon
\end{equation*}
where $\epsilon \sim N(0,1)$ is independent of predictor variables. We can see that the final dataset has only five non-zero $\beta_j$'s, yielding a sparse true vector $\beta$. That is, $\beta_1 = 0.7 \Phi^{-1}(\tau)$ and $\beta_6 = \beta_{12}=\beta_{15}=\beta_{20}=1$, where $\Phi^{-1}(\tau)$ is the $\tau$th quantile of a standard normal distribution. \\

In section \ref{subsec:regpath}, we run QCD and QICD on two different datasets to look at the shape of the relative mean squared error (RMSE) plot along the regularization path. Then, we compare the average runtime of QCD, QICD and LP on multiple scenarios in section \ref{subsec:time} and highlight the computational efficiency of QCD. In section \ref{subsec:accuracy}, we report the estimation error of QCD and demonstrate that it is comparable with QICD and LP.

\subsection{Regularization path}\label{subsec:regpath}

To emphasize the role of warm start and nudge in our pathwise algorithm, we plot the RMSE values while building a regularization path with grid of $\lambda$'s. 
%

The performance metric RMSE is defined as
\begin{equation*}
    RMSE = \frac{\sum_{j=1}^p \left( \hat{\beta}_j - \beta_j \right)^2}{\sum_{j=1}^p \beta_j^2}
\end{equation*}
where $\hat{\beta} = (\hat{\beta_1}, \ldots, \hat{\beta_p})^\top$ is the estimated coefficients and $\beta = (\beta_1, \ldots, \beta_p)^\top$ is the true value. In a pathwise algorithm with $\ell_1$ penalty, we start from $\lambda_{\max}$, where all estimated coefficients become zeros, leading to RMSE of $1$. Once we choose a proper value of $\lambda$, the estimated coefficients will be similar to the true $\beta$ and we would obtain low RMSE value. As we proceed to the next $\lambda$, the RMSE will steadily increase since the estimated $\hat{\beta}$ departs from the truth. \\

\begin{figure}[t!]
\begin{center}
\centerline{\includegraphics[width=\columnwidth]{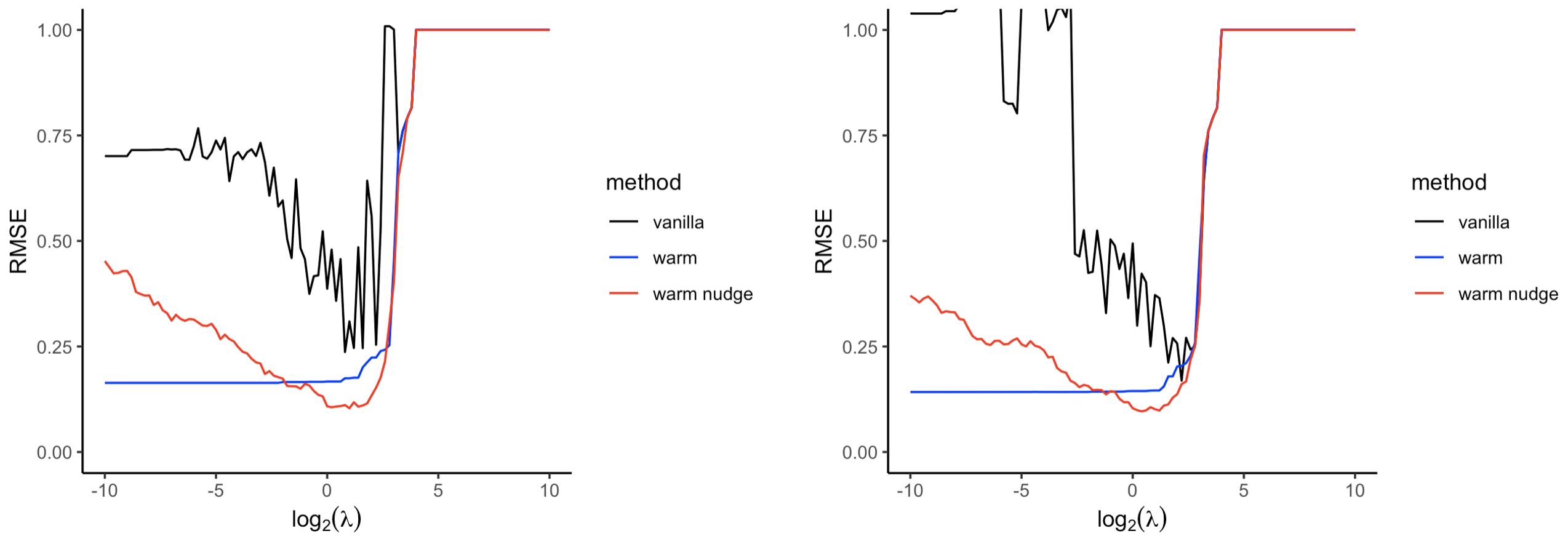}}
\caption{
\textit{Regularization path of vanilla, warm, warm nudge methods per QCD and QICD algorithms in high dimension ($p=150$, $n=50$).} (left) RMSE plot of QCD. (right) RMSE plot of QICD. Adding nudge helps the warm method to converge to an optimal point, yielding lower RMSE value.}
\label{fig:qcd_qicd_high}
\end{center}
\vskip -0.2in
\end{figure}

\begin{table}[t]
\centering
\tabcolsep=19pt
{\small
\resizebox{\textwidth}{!}{%
\begin{tabular}{c|cccc}
\hline
& \multicolumn{4}{c}{min RMSE (\%)}  \\ 
& \multicolumn{2}{c|}{p = 150, n = 50}       & \multicolumn{2}{c}{p = 150, n = 150}  \\
method  & \multicolumn{1}{c}{QCD}                                                        & \multicolumn{1}{c|}{QICD}                                                     & \multicolumn{1}{c}{QCD}                                                      & QICD                                                     \\ 
\hline
\begin{tabular}[c]{@{}c@{}}vanilla\end{tabular}    & \multicolumn{1}{c}{\begin{tabular}[c]{@{}c@{}}18.40 (12.24)\end{tabular}}  & \multicolumn{1}{c|}{\begin{tabular}[c]{@{}c@{}}10.55 (5.86)\end{tabular}} & \multicolumn{1}{c}{\begin{tabular}[c]{@{}c@{}}3.68 (0.25)\end{tabular}} & \begin{tabular}[c]{@{}c@{}}3.79 (0.27)\end{tabular} \\ 
\begin{tabular}[c]{@{}c@{}}warm \end{tabular}       & \multicolumn{1}{c}{\begin{tabular}[c]{@{}c@{}}15.43 (14.24) \end{tabular}} & \multicolumn{1}{c|}{\begin{tabular}[c]{@{}c@{}}7.93 (4.71) \end{tabular}} & \multicolumn{1}{c}{\begin{tabular}[c]{@{}c@{}}3.71 (0.30) \end{tabular}} & \begin{tabular}[c]{@{}c@{}}3.76 (0.26)\end{tabular} \\ 
\begin{tabular}[c]{@{}c@{}}warm nudge\end{tabular} & \multicolumn{1}{c}{\begin{tabular}[c]{@{}c@{}}6.77 (2.63)\end{tabular}}    & \multicolumn{1}{c|}{\begin{tabular}[c]{@{}c@{}}6.00 (1.69)\end{tabular}}  & \multicolumn{1}{c}{\begin{tabular}[c]{@{}c@{}}3.72 (0.27)\end{tabular}}  & \begin{tabular}[c]{@{}c@{}}3.74 (0.28)\end{tabular}  \\ 
\hline
\end{tabular}%
}}
\caption{\textit{Minimum RMSE (in \%) of QCD and QICD for different dimensions $p=150, n=50$ and $p=150, n=150$.} Three methods - vanilla, warm, and warm nudge - were used for each algorithm. Standard deviations (in \%) are reported in parentheses. The results are averaged over 20 different seeds.}
\label{tab:minRMSE_high_low}
\end{table}

Hence, the RMSE plot along the regularization path will display a convex U shape. We will express that the regularization path is stuck if the U shape does not appear in the RMSE plot. 
We generate two dataset with different dimensions: one is low dimension with $p=150, n=150$ and the other is high dimension with $p=150, n=50$. The quantile is set to $\tau=0.3$. We compare three versions of each algorithm QCD and QICD: (i) \textit{vanilla} version of these algorithms do not utilize warm start nor nudge, (ii) \textit{warm} version uses only warm start within the pathwise CD, and (iii) \textit{warm nudge} version implements both warm start and the nudge. The regularization paths of high dimension are reported in Figure \ref{fig:qcd_qicd_high} and those of low dimension are reported in Appendix \ref{apdx:regpath} Figure \ref{fig:qcd_qicd_low}.\\

In Figure \ref{fig:qcd_qicd_high}, we see that the vanilla version for both QCD and QICD display bumpy regularization path. Once warm start is implemented, the regularization path smoothens but the RMSE value maintains the same after a specific $\lambda$. If a nudge is added to warm start, the regularization path starts to display a U shape and we have overcome the stagnation. In particular, the regularization path of warm QCD and warm QICD are stuck in a suboptimal point, yielding higher RMSE than the warm nudge versions. Once we give a nudge, the regularization paths of QCD and QICD display a U-shape with a lower RMSE. This suggests that the nudge plays a crucial role in helping the algorithm converge to an optimal point. Similar patterns are observed in low dimension in Figure \ref{fig:qcd_qicd_low}.\\

The effectiveness of nudge is also confirmed in Table \ref{tab:minRMSE_high_low}. For both QCD and QICD, the warm nudge version's minimum RMSE is much lower compared to the vanilla and warm versions. Note that the QCD in the following sections incorporate warm start and nudge together.

\subsection{Computation time}\label{subsec:time}

\begin{figure}[t]
\begin{center}
\centerline{\includegraphics[width=\columnwidth]{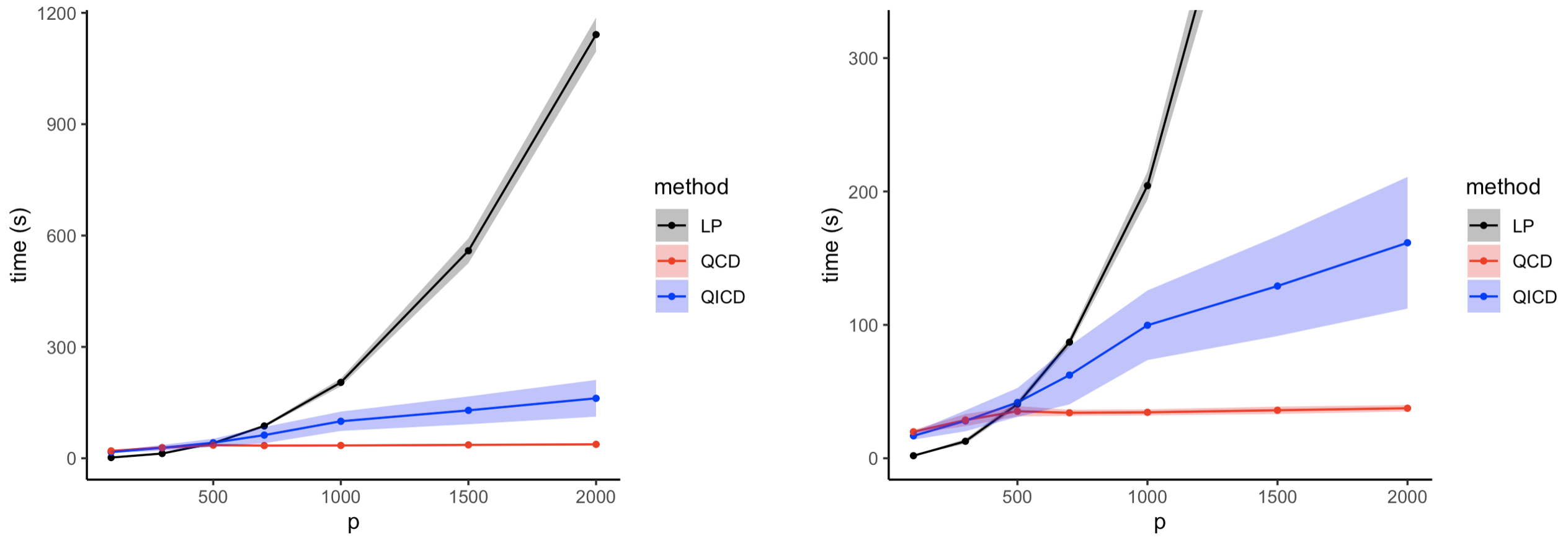}}
\caption{\textit{Average runtime of LP, QICD, and QCD with stopping rule.} (left) Full visualization of the average runtime over 20 different seeds. (right) Enlarged visualization of the average runtime up to 300 seconds. One standard deviation error bars are reported around. The dots correspond to the runtime of each method for each dimension. As dimension grows, QCD displays the lowest computational time compared to LP or QICD.}
\label{fig:runtime_stop}
\end{center}
\vskip -0.2in
\end{figure}

\begin{table}[t]
\tabcolsep=22pt
{\small
\resizebox{\textwidth}{!}{%
\begin{tabular}{c|ccc}
\hline
\multicolumn{1}{l}{} & \multicolumn{3}{|c}{time (seconds)}                \\
dimension (n = 300)           & LP              & QICD            & QCD             \\ \hline
p = 100                & 1.91 (0.06) & 16.83 (2.85)  & 19.86 (1.38) \\ 
p = 300                & 12.74 (1.85) & 28.07 (7.84) & 28.64 (4.55) \\ 
p = 500                & 40.60 (2.50) & 41.83 (10.85) & 35.27 (3.87) \\ 
p = 700                & 87.06 (3.39)  & 62.34 (21.94) & 34.09 (2.31) \\ 
p = 1000               & 204.36 (10.58) & 99.71 (26.08) & 34.45 (2.27) \\ 
p = 1500               & 608.10 (20.57) & 559.05 (33.59) & 36.00 (2.76) \\ 
p = 2000               & 1141.39 (45.84) & 161.57 (49.37) & 37.50 (2.40) \\
\hline
\end{tabular}}%
}
\caption{\textit{Average runtime, in seconds, of LP, QICD, and QCD for different dimensions $p$ when the stopping rule is applied.} The standard deviations of runtime are reported inside the parentheses.}
\label{tab:runtime}
\end{table}

In this subsection, we highlight the computational cost of QCD compared to that of QICD and LP. We generate dataset with dimensions $p=100, 300, 500, 700, 1000, 1500, 2000$ and fixed $n=300$ following the same simulation setting above. Once we achieve the minimum RMSE, the RMSE increases as we decrease the $\lambda$ value and the estimated $\hat{\beta}$ is not sparse anymore, deviating from the truth. Hence, we apply the stopping rule \cite{deb2024regularized} to focus on the area where sparse solution is obtained and solve the convergence issue for small $\lambda$. The stopping rule is
\begin{align}\label{stoprule}
    {RMSE_\lambda - RMSE_{\lambda_{\min}}} > 0.3 \times \left( {RMSE_{\lambda_0}} - {RMSE_{\lambda_{\min}}} \right)
\end{align}
where the $RMSE_{\lambda_0}$ is the RMSE when $\lambda_{\max}$ is applied, $RMSE_{\lambda}$ is the RMSE value of the current $\lambda$, and $RMSE_{\lambda_{\min}}$ is the minimum RMSE. In other words, we use this rule of thumb to stop computing solutions for clearly suboptimal values of $\lambda$. While this rule requires knowledge of ground truth, similar rules can be designed in real data sets by changing the RMSE to AIC or BIC scores.\\

We create 20 replicates for each dimension $p$ to report the average runtime and its standard deviation when stopping rule is applied for each method in seconds in Figure \ref{fig:runtime_stop} and Table \ref{tab:runtime}. Figure and table for the average runtime when stopping rule is not applied are given in Appendix \ref{apdx:comptime} Figure \ref{fig:runtime_nostop} and Table \ref{tab:runtime_nostop}.\\

In Figure \ref{fig:runtime_stop}, we can see that as the dimension of the dataset increases, the computation time of LP increases linearly, up to 1000 seconds at the end. To compare the runtime of QICD and QCD, we focus on the zone within 300 seconds in the right plot. In low dimension $p=100$ and $300$, QCD has slightly lower computational speed than QICD or LP. However, once we go to high dimension where $p>300$, QCD's speed remains nearly constant compared to QICD or LP and becomes much faster. The reason for this boost is due to the KKT condition check within the Algorithm \ref{alg1}. In a much higher dimension, the estimated $\hat{\beta}$ is sparser than the $\hat{\beta}$ of low dimension. Since the KKT condition is checking exactly $\hat{\beta} = 0$, there will be more cases where we do not sort our data points in high dimension, gaining computational benefit. We also observe that the error bar of QICD is larger than the error bar of QCD or LP and it widens as dimension grows. This means that the regularization paths of QICD are much more wiggly, approximately 10 to 20 times more volatile than those of QCD as listed in Table \ref{tab:runtime}. This emphasizes the merit of warm start and nudge implemented in our QCD.

\subsection{Estimation and model selection accuracy}\label{subsec:accuracy}

\begin{table}[t!]
\centering
\tabcolsep=27pt
{\small
\resizebox{\textwidth}{!}{%
\begin{tabular}{c|ccc}
\hline
\multicolumn{1}{l}{} & \multicolumn{3}{|c}{min RMSE}                \\
dimension (n = 300)           & LP              & QICD            & QCD             \\ \hline
p = 100                & 3.44 (0.12)           & 3.44 (0.12)           & 3.44 (0.12) \\ 
p = 300                & 3.48 (0.15)  & 3.50 (0.16)           & 3.49 (0.15)         \\ 
p = 500                & 3.58 (0.21)           & 3.59 (0.21)           & 3.57 (0.18) \\ 
p = 700                & 3.50 (0.21)           & 3.50 (0.21)           & 3.51 (0.19) \\ 
p = 1000              & 3.56 (0.07)           & 3.57 (0.07)           & 3.55 (0.07) \\ 
p = 1500              & 3.61 (0.19)           & 3.63 (0.20)           & 3.62 (0.15) \\ 
p = 2000              & 3.64 (0.19)           & 3.67 (0.22)           & 3.65 (0.19) \\
\hline
\end{tabular}}%
}
\caption{\textit{Minimum RMSE (in \%) of LP, QICD, and QCD for different dimensions.} Standard deviations (in \%) are reported in parentheses. The results are averaged over 20 different seeds.}
\label{tab:minRMSE}
\end{table}

\begin{table}[t!]
\centering
\tabcolsep=27pt
{\small
\resizebox{\textwidth}{!}{%
\begin{tabular}{c|ccc}
\hline
\multicolumn{1}{l}{} & \multicolumn{3}{|c}{AUROC (\%)}                \\
dimension (n = 300)           & LP              & QICD            & QCD             \\ \hline
p = 100                & 89.18 (4.54) & 89.17 (3.53)  & 89.18 (2.67) \\ 
p = 300                & 88.95 (5.78) & 89.38 (3.73) & 89.11 (4.36) \\ 
p = 500                & 88.67 (5.28) & 90.43 (2.20) & 89.63 (2.13) \\
p = 700                & 88.87 (6.07)  & 88.96 (4.22) & 89.57 (3.76) \\ 
p = 1000               & 91.28 (4.20) & 89.99 (0.08) & 90.09 (0.11) \\ 
p = 1500               & 87.95 (5.89) & 89.99 (0.04) & 89.13 (2.92) \\ 
p = 2000               & 86.65 (5.97) & 89.98 (0.04) & 89.56 (2.21) \\ \hline
\end{tabular}}%
}
\caption{\textit{AUROC (in \%) of LP, QICD, and QCD for different dimensions.} Standard deviations (in \%) are reported in parentheses. AUROC values are those that correspond to the $\log_2(\lambda)$ value when minimum RMSE is achieved. The results are averaged over 20 different seeds.}
\label{tab:AUROC}
\end{table}

After demonstrating the computational benefit of QCD, we assess the accuracy of QCD with the same dataset from  section \ref{subsec:time}. We use RMSE and area under the ROC curve (AUROC) as the performance metric and compare QCD with LP and QICD. The RMSE calculates elementwise difference between the estimated coefficients and the true coefficients. AUROC, on the other hand, is an indicator of whether the sparsity pattern is correctly recovered. 
\\

We report the average minimum RMSE (in \%) of LP, QICD, and QCD with the standard deviations (in \%) in parentheses in Table \ref{tab:minRMSE}. The reported values are averaged over 20 different seeds of data. For every dimension, the minimum RMSE values for QCD are very close to the minimum RMSEs of LP or QICD. \\

We then construct Table \ref{tab:AUROC} which shows the AUROC values and standard deviations (in \%) of LP, QICD, and QCD when $\log_2(\lambda)$ corresponding to the minimum RMSE are used for each dimension. We notice that the AUROC rates of QCD is similar to that of LP or QICD in every dimension. This suggests that QCD recovers the sparsity pattern of the true coefficients correctly as LP and QICD. Further visualization and interpretation of the RMSE and AUROC plots for all seven dimensions are displayed in Appendix \ref{apdx:acc}. 

%% file: 05-conclusion.tex
\section{Conclusion}\label{sec:conclusion}

Penalized quantile regression is a useful algorithm for high-dimensional data analysis, but its applications in large-scale problems is somewhat limited due to computational bottlenecks stemming from the nonsmooth nature of the problem. We developed a fast pathwise coordinate descent algorithm for computing the exact $\ell_1$-penalized quantile regression. Two core features of our algorithm, viz. the strategy of derivative tracking for nonsmooth loss functions with piecewise constant derivatives, and random perturbation to help algorithm from getting stuck along the regularization path, are new contributions to the literature and hold promise to be applicable more broadly. A FORTRAN 77 implementation of our algorithm is also provided in the form of an R package, and can be used as the core to build Python or MATLAB packages as well. \\

The promising empirical performance of our algorithm on simulated data motivates several questions. On the algorithmic side, the amount of nudge can be made data-adaptively to increase speed for non-sparse solutions. On the computational side, the algorithm can benefit from faster ways to compute the weighted median with partial sorting of the data. On the theoretical side, a full convergence analysis of the coordinate descent algorithm with nonsmooth, nonseparable check loss function can help assess the scope of the proposed algorithm. We leave these for future research directions.

\section{Acknowledgements}

SK and SB acknowledge partial support from NIH award R01GM135926. In addition, SB
acknowledges partial support from NSF awards DMS-1812128, DMS-2210675, and CAREER DMS-2239102. SK and SB thank Dr. Jonas Krampe for helping on the illustration of the exact and approximate coordinate descent updates.

%% file: 06-appendix.tex
\section*{Appendix}

\section{Additional Background Details}\label{apdx:background}

This section contains more detailed description of Section \ref{sec:background}. We explain the computational aspects of quantile regression, both in classical and high-dimensional, penalized setups. 

\subsection{Quantile Regression}

The $\tau$th sample quantile defined in \citet{koenker1978regression} is
\begin{equation*}
    F^{-1}(\tau) = \inf\{ y : F(y) \geq \tau \}
\end{equation*}
$\forall$ $\tau \in (0,1)$ where $F(y)$ is the cumulative distribution function of $Y$. This sample quantile is the solution to minimization the problem 
\begin{equation}
     \argmin_{c \in \mathbb{R}} L(c) = \argmin_{c \in \mathbb{R}} \sum_{i=1}^n \rho_\tau (y_i - c).
\tag{\ref{eqn:qr}}
\end{equation}
This is the simplest qauntile finding problem since it only involves one dimensional $y_i$ vector. In order to find the solution of \eqref{eqn:qr}, let us take a look at the derivative $L'(c)$ following the definition of check-loss function $\rho_\tau \left(u\right):= u \left(\tau-\mathbf{I}(u<0)\right)$. Note that the solution will occur when $L'(c)=0$ since $L(c)$ is convex.
\begin{equation}
    L'(c) = \sum_{i \in \{ i : y_i \geq c \} } (-\tau) + \sum_{i \in \{ i : y_i < c \} } (1-\tau) 
\tag{\ref{eqn:qr_Lprime}}
\end{equation}
Then, this yields a solution of $\tau\{\#i : y_i \geq c \} = (1-\tau)\{\#i : y_i < c \}$. Solving \eqref{eqn:qr} will only require one round of sorting through the scalar data set since we only need to compare the magnitude of $y_i$ and $c$. For example, when $\tau=0.5$, the estimator is called the sample median and the problem becomes the least absolute deviation (LAD) where the solution occurs when $\{\#i : y_i \geq c \} = \{\#i : y_i < c \}$. This median finding is equivalent to finding the zero crossing of the derivatives $L'(c)$. We display the objective function $L(c)$ and the derivative $L'(c)$ plots for $\tau=0.5$ in Figure \ref{fig:simpleQR}. We observe that $L(c)$ is piecewise linear and $L'(c)$ is piecewise constant, monotone increasing. Hence, the solution $\hat{c}$ in \eqref{eqn:qr} will occur when $L'(c)$ first crosses zero. In addition, the difference between each slope is always $\pm 1$ for any $\tau$.

\subsection{Penalized Quantile Regression}\label{subsec:pqr}

We can expand the simplest quantile finding problem into a higher dimension $\mathbb{R}^p$ with response and predictor variables. 
\begin{equation} 
    \min_{\boldsymbol{\beta} \in \mathbb{R}^p} \ \sum_{i=1}^n \rho_\tau (y_i - x_i^\top \beta). 
\tag{\ref{eqn:qr-pdim}}
\end{equation}
$\rho_\tau(\cdot)$ is a check loss function as mentioned above, $x_i = (x_{i1}, \ldots, x_{ip})^\top$, and $\beta = (\beta_1, \ldots, \beta_p)^\top$.
And by adding a penalty, we can write the general penalized QR problem as
\begin{equation}
    \min_{\boldsymbol{\beta} \in \mathbb{R}^p} \ \sum_{i=1}^n \rho_\tau (y_i - x_i^\top \beta) + \sum_{j=1}^p p_\lambda(|\beta_j|) 
\tag{\ref{eqn:pqr}}
\end{equation}
where $p_\lambda(\beta_j)$ is the penalty function of our choice, and $\lambda$ is the penalty. By choosing an appropriate penalty, we can perform variable selection that yields a sparse solution. The choice of penalty function $p_\lambda(\beta)$ can vary whether we use a convex or a non-convex penalty. \\

\citet{tibshirani1996regression} introduced the convex Lasso penalty ($\ell_1$ penalty), which is a sum of the absolute value of coefficients. $\ell_1$ penalty makes some coefficients exactly to zero due to its formula and produce a stable and concise model. In this work, we develop an optimization algorithm for the $\ell_1$-QR problem, which is \eqref{eqn:pqr} when $p_\lambda(|\beta_j|)=\lambda|\beta_j|$. For $\tau=0.5$, the penalized QR is called LAD-Lasso and the properties of the estimator is described in \citet{wang2007robust}. In the high dimension setting where $p$ is very large, \citet{belloni2011ℓ} studied the properties of $\ell_1$-QR estimators under the sparsity assumption, when the amount of non-zero coefficients are less than the sample size $n$. Apart from the convex penalty Lasso, one could use non-convex penalties. The smoothly clipped absolute deviation (SCAD) proposed by \citet{fan2001variable} and minimax concave penalty (MCP) found by \citet{zhang2010nearly} are those examples.

\subsection{Linear Programming}\label{subsec:lp}

The most widely used methods to solve the QR problem are simplex and interior-point method. Both methods require reformulating \eqref{eqn:qr-pdim} into a linear program as below.
\begin{align} \label{eqn:lp}
    &\min_{\boldsymbol{\beta}} \sum_{i=1}^n ( \tau u_i +(1-\tau)v_i ) \nonumber \\
    &\text{subject to} \nonumber \\
    & -v_i \leq y_i - x_i^T \beta \leq u_i \nonumber \\
     &u_i \geq 0, \ v_i \geq 0, \ i = 1,2, \ldots, p. 
\end{align}
Here, the slack variables $u_i$ and $v_i$ are positive and negative parts of $y_i - x_i^T \beta$ respectively.
Then, for simplex method, one can follow the modified algorithm of \citet{barrodale1974algorithm} that \citet{koenker2005quantile} provides to obtain the optimal solution. Geometrically, one can understand that simplex method moves along the edges of the formed polytope until it reaches the vertex of the optimal solution. \\

For larger sample sizes, interior-point method called Frisch–Newton method suggested by \citet{koenker1996quantile} is preferred due to its linearly growing computation time compared to simplex method's quadratic growth. As we can see from its name, interior-point method starts searching points from the interior of the feasible region and reaches the boundary to find the optimal solution. It utilizes a logarithmic barrier function and combine it to \eqref{eqn:lp} through the Lagrange multiplier, then solve by lessening the duality gap. Detailed solution steps could be found in \citet{koenker2005quantile}. In a presence of a $\ell_1$ penalty as in \eqref{eqn:qr-lasso}, one more convex condition $\sum_{j=1}^p |\beta_j| \leq \lambda$ will be added to \eqref{eqn:lp} and the problem will maintain the structure of linear program, where interior-point methods is applicable. \\

In terms of computation time, \citet{koenker1996quantile} discuss the computational complexity of LP when it is used on median regression. In an iid Gaussian simulated data with large number of observations, the interior-point method was faster than the simplex method as the dimension increased. However, as commented in \citet{he2023smoothed}, in large scale datasets where $p$ becomes as large as $n$ or larger than $n$, the interior-point method becomes very slow or even run out of memory. 

\subsection{Coordinate Descent}\label{subsec:cd}

In addition to the linear programming methods, coordinate descent (CD) is another technique to solve both low and high-dimension penalized problems. CD has been studied on various loss functions including the least squares, LAD, and QR. In this subsection, we first explain how the CD algorithm is applied to each problem without penalty term, and then attach the $\ell_1$ penalty. \\

Consider a least squares problem without penalty first for simplicity. 
\begin{equation} \label{eqn:cd-ls}
    \min_{\beta_j} \ \sum_{i=1}^n \left( y_i - \sum_{k \neq j} x_{ik}\beta_k - x_{ij}\beta_j \right)^2
\end{equation}
The main idea of coordinate-wise algorithm, known as cyclic CD, is to cycle through each $\beta_j$, $j=1,2,\ldots,p$ by fixing the partial residuals $y_i-\sum_{k \neq j} x_{ik} \beta_k $ until convergence for a given penalty $\lambda$. In this problem, the CD solution would be obtained by taking the derivative with respect to $\beta_j$ due to quadratic loss function. Adding $\ell_1$ penalty to \eqref{eqn:cd-ls} will change the solution into a soft-threshold between the partial residuals and penalty $\lambda$ as \citet{friedman2007pathwise} showed.  \\

For the LAD loss function, \citet{edgeworth1887observations} first proposed that the CD solution to LAD problem is equivalent to a weighted median. Then, \citet{li2004maximum} provided details on finding the weighted median in a simple setting $\sum_{i=1}^n |y_i - a x_i - b|$. If we apply the same logic to the $p$-dimension LAD problem, we can express the problem as
\begin{equation} \label{eqn:cd-lad}
    \min_{\beta_j} \ \sum_{i=1}^n |x_{ij}| \left| \frac{y_i - \sum_{k \neq j} x_{ik}\beta_k}{x_{ij}} - \beta_j \right|.
\end{equation}
By factoring $|x_{ij}|$ out, we can view $(y_i - \sum_{k \neq j} x_{ik}\beta_k)/x_{ij}$ as  modified partial residuals and consider them as fixed. Without loss of generality, $x_{ij} \neq 0$ since the summand would not involve the minimizing variable $\beta_j$ when $x_{ij}=0$. We could see that iterating for each $\beta_j$ in \eqref{eqn:cd-lad} only requires one round of sorting through the scalar dataset of modified partial residuals. This procedure is eventually reduced into finding a weighted median of the modified partial residuals where $|x_{ij}|$ is the weight. 
In addition, \citet{wu2008coordinate} proposed a greedy version of CD to solve \eqref{eqn:cd-lad} based on Edgeworth's algorithm and concluded that it is faster than cyclic CD. Their CD is greedy in terms of updating $\beta_j$. They calculate the forward and backward directional derivatives for all $\{ \beta_j \}_{j=1,\ldots,p}$ at once and update $\beta_j$ that has the steepest decrease in the objective function.\citet{wu2008coordinate} also stated that adding $\ell_1$ penalty to \eqref{eqn:cd-lad} is same as adding partial residual of zero and weight of $\lambda$. Considering the penalty term as an additional data point is going to be a key scheme when it comes to applying CD to $\ell_1$-QR. We also utilize this idea in our algorithm.\\

The QICD algorithm proposed by \citet{peng2015iterative} solves non-convex penalized QR problem through iterative CD. Even the paper only describes the CD algorithm for SCAD, MCP penalized QR, we can easily apply the same logic to $\ell_1$-QR problem. Furthermore, they provide users an option to select Lasso penalty in their $\texttt{R}$ package $\texttt{QICD}$. We now describe their QICD algorithm to solve $\ell_1$-QR after carefully reviewing the logic and the package. When iterating until convergence to obtain a set of vector $\left( \hat{\beta_1}, \ldots, \hat{\beta_p} \right)$, the estimated $\hat{\beta_j}^{(r+1)}$ at subiteration $r+1$ can be written as 
\begin{align} 
    \hat{\beta_j}^{(r+1)} &= \argmin_{\beta_j} \left\{ \frac{1}{n} \left[ \sum_{i=1}^n \rho_\tau \left(y_i - \sum_{s<j} x_{is} \hat{\beta_s}^{(r+1)} - \sum_{s>j} x_{is} \hat{\beta_s}^{(r)} - x_{ij} \beta_j) \right) \right] + \lambda \sum_{s<j}|\hat{\beta_s}^{(r+1)}| + \lambda \sum_{s>j}|\hat{\beta_s}^{(r)}| + \lambda|\beta_j| \right\} \nonumber \\
    &= \argmin_{\beta_j} \left\{ \frac{1}{n} \sum_{i=1}^n \rho_\tau    \left( x_{ij}u_{ij} \right)  + \lambda \sum_{s<j}|\hat{\beta_s}^{(r+1)}| + \lambda \sum_{s>j}|\hat{\beta_s}^{(r)}| + \lambda|\beta_j| \right\} \nonumber \\
    &= \argmin_{\beta_j} \left\{ \frac{1}{n} \sum_{i=1}^n \left| x_{ij} u_{ij}  \left( \tau - \mathbf{I}(x_{ij} u_{ij} < 0) \right) \right|  + \lambda \sum_{s<j}|\hat{\beta_s}^{(r+1)}| + \lambda \sum_{s>j}|\hat{\beta_s}^{(r)}| + \lambda|\beta_j| \right\} \nonumber \\
    &= \argmin_{\beta_j} \left\{ \sum_{i=1}^n \frac{1}{n} \left| x_{ij}   \left( \tau - \mathbf{I}(x_{ij} u_{ij} < 0) \right) \right| |u_{ij}|   + \lambda \sum_{s<j}|\hat{\beta_s}^{(r+1)}| + \lambda \sum_{s>j}|\hat{\beta_s}^{(r)}| + \lambda|\beta_j| \right\} \nonumber \\
    &:= \argmin_{\beta_j} \left\{ \frac{1}{n+1} \sum_{i=1}^{n+1} w_{ij} |u_{ij}| \right\} 
    \tag{\ref{eqn:qicd}}
\end{align}
where
\begin{align*}
   &u_{ij} = \begin{dcases}
			\frac{y_i - \sum_{s<j} x_{is} \hat{\beta_s}^{(r+1)} - \sum_{s>j} x_{is} \hat{\beta_s}^{(r)}}{x_{ij}} - \beta_j, & i=1, \ldots, n \\
            \beta_j, & =n+1,
		 \end{dcases}  \\
    &w_{ij} = \begin{dcases}
    \frac{1}{n} \left| x_{ij} \left( \tau - \mathbf{I}(x_{ij} u_{ij} < 0) \right) \right|, & i=1, \ldots, n \\
    \lambda, & i=n+1.
    \end{dcases}
\end{align*}
The number of observations increased from $n$ to $n+1$ where an additional pseudo zero partial residual is added, so we can view $|\beta_j| = |0-\beta_j|$. The important thing to note here is that QICD is not giving the exact CD solution. This is because they have $\beta_j$ inside the weights $w_{ij}$, which should be considered as a fixed value. Hence, we solve \eqref{eqn:qicd} exactly and derive an exact CD update in Section \ref{sec:method}. See \ref{sec:mcp} and \ref{sec:scad} for the exact CD update of SCAD and MCP penalized QR. 

\subsection{Pathwise Coordinate Descent}\label{subsec:pathwisecd}

Apart from estimating coefficients through $\ell_1$-QR with one $\lambda$, we can consider a grid of $\lambda$ ranging from $\lambda_{min}$ close to zero to $\lambda_{max}$. The key is to start from $\lambda_{\max}$ and decrease the $\lambda$ value gradually. Combining this scheme with CD, we can repeatedly run CD for each $\lambda$ value within the grid and explicitly view how the regularization path is constructed.\\

While varying $\lambda$ values, a technique called \textit{warm start} is performed to lower the computation time. \citet{friedman2010regularization} showed how warm start is applied to $\ell_1$ squared loss problem and detailed description of warm start is contained in \citet{hastie2015statistical}.  Furthermore, The computational benefit of pathwise CD was shown in \citet{friedman2007pathwise} for Lasso penalty. We first start from $\lambda_{\max}$ within the grid of $\lambda_{\ell} \in [\lambda_{\min}, \lambda_{\max}], \ {\ell= 1, \ldots, L}$, where $L$ is the total number of $\lambda$ values within the grid. It uses the previous solution $\hat{\beta}(\lambda_{\ell-1})$ as a warm start for estimating next $\hat{\beta}(\lambda_{\ell})$. This implies that we are starting from the coefficient values that are close to the real ones, so the non-zero coefficients will be recovered rapidly and fewer iteration will be required for each $\lambda_\ell$. As we start from $\lambda_{max}$, the initial coefficients will all be zeros and as $\lambda$ decreases, the non-zero coefficients will show up towards the end of the solution path. \\

Another scheme that could be applied within the \textit{pathwise} algorithm is active set selection. In this work, we only implement warm start but not active set selection. At step $\ell$ with $\lambda_\ell$, we only store the non-zero coefficient set (active set) $\mathcal{A}_{\ell-1}$. Then, we check whether each coefficient that are not included in $\mathcal{A}_{\ell-1}$ satisfies the simple pass fail test. For \eqref{eqn:cd-ls} with $\ell_1$ penalty, the test is $\frac{1}{N}|\langle x_j, r \rangle| < \lambda_\ell$. Here, $x_j$ is $(x_{1j}, \ldots, x_{nj})$ and $r$ is the current partial residuals. The coefficients that did not pass the test are included into $\mathcal{A}_{\ell-1}$ and we repeat the pathwise CD. \citet{hastie2015statistical} showed that active set selection dramatically boosts the speed of pathwise CD.

\section{Theoretical Results for QCD algorithm}\label{apdx:theory}

In this section, we present the proof of Proposition \ref{prop1} and \ref{prop3} in \ref{apdx:derivative_proof} and provide a convergence simulation of QCD and compare it with QICD in \ref{apdx:qcd_conv}.

\subsection{Proofs}\label{apdx:derivative_proof}

\noindent \textbf{Proposition 3.1.}
\textit{Define} 
$S' = -\sum_{i=1}^{n} |x_i|  \left( \tau \mathbf{I}(x_i > 0) + (1-\tau)\mathbf{I}(x_i < 0)  \right)$. \textit{Then, the derivative when} $v_m < \beta < v_{m+1} < \ldots < 0$ \textit{can be expressed as} 
$$S_m' = S' + \sum_{i=1}^m |x_i| - \lambda.$$
\textit{The derivative when} $\beta$ \textit{exceeds} $0$, $0 < \ldots <v_q < \beta < v_{q+1}$, \textit{can be expressed as}
$$S_{q}' = S' + \sum_{i=1}^q |x_i| + \lambda.$$

\begin{proof}
We start from the first range, $\beta<v_1$. We know that $\beta<0$ and every $v_i-\beta$ in \eqref{eqn:obj_qrlasso_simple} will be positive due to the range we defined. Based on the definition of $\rho$, $\rho_\tau(v_i-\beta) = \tau(v_i-\beta)$ and $\rho_{1-\tau}(v_i - \beta) = (1-\tau)(v_i-\beta)$. Then, we can explicitly write the objective function value $S_0$ and its derivative $S_0'$ as below.
\begin{align*}
    \bullet \ &\beta < v_1 \\
    S_0 &= \sum_{i : x_i>0} |x_i| \tau (v_i-\beta) + \sum_{i : x_i<0} |x_i| (1-\tau) (v_i-\beta) - \lambda \beta \\ &= \sum_{i=1}^{n} |x_i|  \left( \tau \mathbf{I}(x_i > 0) + (1-\tau)\mathbf{I}(x_i < 0)  \right) (v_i-\beta) - \lambda \beta \\ &= S - \lambda \beta \\
    S_0' &= -\sum_{i : x_i>0} |x_i| \tau -\sum_{i : x_i<0} |x_i| (1-\tau) - \lambda \\ &= -\sum_{i=1}^{n} |x_i|  \left( \tau \mathbf{I}(x_i > 0) + (1-\tau)\mathbf{I}(x_i < 0)  \right)- \lambda \\ &= S'-\lambda
\end{align*}
We proceed to the next range where $v_1<\beta<v_2$. Within this range, only $v_1-\beta$ will be negative, so the check loss values in \eqref{eqn:obj_qrlasso_simple} are either $\rho_\tau(v_1-\beta) = (\tau-1)(v_1-\beta)$ or $\rho_{1-\tau}(v_1-\beta) = -\tau(v_1-\beta)$. An important fact to note here is that the objective function and its derivative are the same regardless of the range of $x_i$. That is, one does not need to consider two separate cases, whether $x_1>0$ or $x_1<0$. We could easily verify that this is true from below calculation. 
We subtract the already included $v_1-\beta$ term from $S_0$ and add a newly calculated $v_1-\beta$ term for both ranges $x_1>0$ and $x_1<0$. 
\begin{align*}
    \bullet \ v_1 &< \beta < v_2 \\
    (i) \ \ &x_1>0 & (ii) \ \ &x_1<0 \\
    S_1 &= |x_1| (\tau-1)(v_1-\beta) + S - |x_1|\tau(v_1-\beta) -\lambda \beta & S_1 &= |x_1| (-\tau)(v_1-\beta) + S- |x_1|(1-\tau)(v_1-\beta) -\lambda \beta \\  &= S - \lambda\beta - |x_1|(v_1-\beta) &  &= S - \lambda\beta - |x_1|(v_1-\beta) \\ &= S_0 - |x_1|(v_1-\beta) & &= S_0 - |x_1|(v_1-\beta) \\
    S_1' &= S'-\lambda + |x_1|  & S_1' &= S' - \lambda + |x_1|  \\ &= S_0' + |x_1| & &= S_0' + |x_1|
\end{align*}

Following the same logic, this recursive form continues until the range $v_1< v_2 < \ldots < v_{k} < \beta < 0$. $v_1-\beta, v_2-\beta, \ldots, v_k-\beta$ are all negative and we do not need to consider the range of $x_1, x_2, \ldots, x_k$. Then, the $k$th objective function and its derivative are
\begin{align*}
    \bullet \ &v_1 < \ldots < v_k < \beta < 0 \\
     S_{k} &= S - \sum_{i=1}^k |x_i| (v_i-\beta) -\lambda \beta \\ &= S_{k-1} -  |x_{k}| (v_{k}-\beta) \\
     S_{k}' &= S' + \sum_{i=1}^k |x_i|-\lambda \\  &= S_{k-1}' + |x_{k}|.
\end{align*}

Once $\beta$ exceeds $0$, $\beta>0$ will lead $\lambda |\beta| = \lambda \beta$ in \eqref{eqn:obj_qrlasso_simple}. The first time when this happens is range $v_1< v_2 < \ldots < v_k < 0 < \beta < v_{k+2}$. From this range, we need to add $2\lambda$ to the recursive form.  Hence, the $k+1$th objective function and its derivative are
\begin{align*}
    \bullet \ v_1 &< \ldots < v_k < 0 < \beta < v_{k+2} \\
     S_{k+1} &= S  - \sum_{i=1}^{k+1} |x_i| (v_i-\beta)+ \lambda \beta \\ &= S_{k} -  |x_{k+1}| (v_{k+1}-\beta) + 2\lambda\beta \\
     S_{k+1}' &= S'  + \sum_{i=1}^{k+1} |x_i| +\lambda \\  &= S_{k}' + |x_{k+1}| +2\lambda.
\end{align*}
The final range is $v_1< \ldots 0 < \ldots < v_{n+1} < \beta$ where
\begin{align*}
    \bullet \ &v_1 < \ldots < v_{n+1} < \beta \\
     S_{n+1} &= S_{n} -  |x_{n+1}| (v_{n+1}-\beta) \\
     S_{n+1}' &= S_{n}' + |x_{n+1}|.
\end{align*}
\end{proof}

\textbf{Proposition 3.3.} \textit{The minimizer of} \eqref{eqn:obj_qrlasso_simple} \textit{is zero if and only if}  
    $$\left|S' + \sum_{i : v_{i}<0} |x_{i}| \right| \leq \lambda.$$ 

\begin{proof} 
Define the ordered $v_i$'s as $v_1<v_2< \ldots < v_k < 0 < v_{k+2} < \ldots < v_{n+1}$. The expression $S' + \sum_{i : v_{i}<0} |x_{i}|$ is equivalent to the derivative when $v_k < \beta < 0$ which is negative. If $\left|S' + \sum_{i : v_{i}<0} |x_{i}| \right| \leq \lambda$, the next derivative will cross $0$ and the solution of \eqref{eqn:obj_qrlasso_simple} will occur at $0$. This is because adding the $\lambda$ boost will make the derivative become larger than $0$. \\

If minimizer of \eqref{eqn:obj_qrlasso_simple} is zero, the derivative crosses zero exactly at $\beta = 0$. So the derivative in the segment right before 0 is negative, and the derivative in the segment right after 0 is positive. This means

$$
S' + \sum_{i: v_i < 0} |x_i| - \lambda < 0, ~~~
S' + \sum_{i: v_i < 0} |x_i| + \lambda \geq 0.
$$

Combining the two inequalities, we can say
$$
 -\lambda \leq S' + \sum_{i: v_i < 0} |x_i| < \lambda.
$$
\end{proof}

\subsection{Exact vs. approximate CD updates}\label{apdx:qcd_conv}


In this subsection, we show that the objective function value of \eqref{eqn:qicd} when solved exactly is different from the value obtained by using QICD with a two dimensional example ($j=1,2$) and suggest QCD yields the coordinatewise minimum.
\begin{equation*}
    L(\beta_1, \beta_2) = \frac{1}{n+1} \sum_{i=1}^{n+1} w_{ij}(\beta_1,\beta_2) |u_{ij}(\beta_1,\beta_2)|
\end{equation*}
Here, $u_{ij}(\beta_1, \beta_2)$ and $w_{ij}(\beta_1, \beta_2)$ will have different formulas for $j=1, 2$. 

For $j=1$,
\begin{align*}
   &u_{i1}(\beta_1,\beta_2) = \begin{dcases}
			\frac{y_i - \sum_{k=1}^n x_{k2}\beta_2}{x_{i1}} - \beta_1, & i=1, \ldots, n \\
            \beta_1, & i=n+1,
		 \end{dcases}  \\
    &w_{i1}(\beta_1, \beta_2) = \begin{dcases}
    \frac{1}{n} \left| x_{i1} \left( \tau - \mathbf{I}(x_{i1} u_{i1} < 0) \right) \right|, & i=1, \ldots, n \\
    \lambda, & i=n+1.
    \end{dcases}
\end{align*}
For $j=2$,
\begin{align*}
    &u_{i2}(\beta_1,\beta_2) = \begin{dcases}
			\frac{y_i - \sum_{k=1}^n x_{k1}\beta_1}{x_{i2}} - \beta_2, & i=1, \ldots, n \\
            \beta_2, & i=n+1,
		 \end{dcases}  \\
    &w_{i2}(\beta_1, \beta_2) = \begin{dcases}
    \frac{1}{n} \left| x_{i2} \left( \tau - \mathbf{I}(x_{i2} u_{i2} < 0) \right) \right|, & i=1, \ldots, n \\
    \lambda, & i=n+1. 
    \end{dcases}
\end{align*}

\begin{figure}[t!]
\vskip 0.2in
\begin{center}
\centerline{\includegraphics[width=\columnwidth]{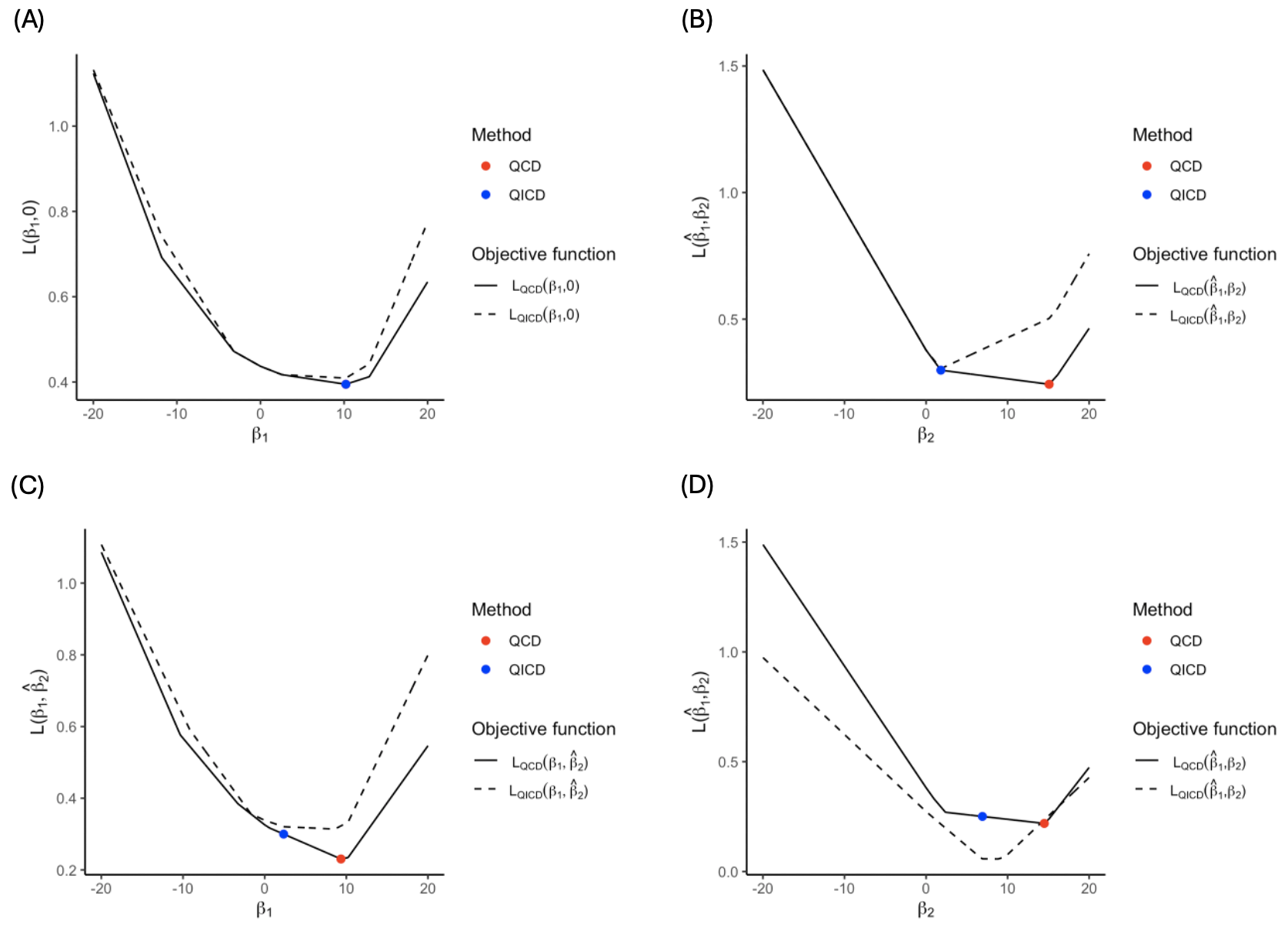}}
\caption{\textit{Objective function values and estimated $\hat{\beta}_1, \hat{\beta}_2$ of the first two iterations of coordinate descent algorithm.} (A) Estimation of $\hat{\beta}_1$ and $L(\beta_1, \beta_2)$ in the first iteration by initializing $\beta_2=0$. (B) Estimation of $\hat{\beta}_2$ and $L(\hat{\beta}_1, \beta_2)$ in the first iteration using $\hat{\beta}_1$ from (A). (C) Estimation of $\hat{\beta}_1$ and $L(\beta_1, \hat{\beta}_2)$ in the second iteration by using $\hat{\beta}_2$ from (B). (D) Estimation of $\hat{\beta}_2$ and $L(\hat{\beta}_1, \beta_2)$ in the second iteration using $\hat{\beta}_1$ from (C). The solid line and the red dot correspond to the objective function and estimated $\hat{\beta}_j$ when our algorithm QCD is used. The dashed line and the blue dot are the objective function and estimated $\hat{\beta}_j$ when QICD is used. For comparison, the blue dot is drawn on top of the solid line.}
\label{fig:converge1}
\end{center}
\vskip -0.2in
\end{figure}

The difference between QCD and QICD is the way we treat $u_{ij}(\beta_1, \beta_2)$. QICD considers $w_{ij}(\beta_1, \beta_2)$ as known by replacing unknown $u_{ij}(\beta_1, \beta_2)$ with $\hat{u}_{ij}$. $\hat{u}_{ij}(\beta_1, \beta_2)$ is obtained through substituting unknown $\beta_j$ with $\hat{\beta}_j$ that was estimated in the previous iteration. So for $j=1$, $u_{i1}(\beta_1, \beta_2)$ and $w_{i1}(\beta_1, \beta_2)$ for QICD could be defined as
\begin{align*}
   &u_{i1}(\beta_1,\hat{\beta}_2) = \begin{dcases}
			\frac{y_i - \sum_{k=1}^n x_{k2}\hat{\beta}_2}{x_{i1}} - \beta_1, & i=1, \ldots, n \\
            \beta_1, & i=n+1,
		 \end{dcases}  \\
       &\hat{u}_{i1}(\beta_1, \hat{\beta}_2) =\frac{y_i - \sum_{k=1}^n x_{k2}\hat{\beta}_2}{x_{i1}} - \hat{\beta}_1 \\
    &w_{i1}(\beta_1, \hat{\beta}_2) = \begin{dcases}
    \frac{1}{n} \left| x_{i1} \left( \tau - \mathbf{I}(x_{i1} \hat{u}_{i1} < 0) \right) \right|, & i=1, \ldots, n \\
    \lambda, & i=n+1. 
    \end{dcases}
\end{align*}
We could apply the same logic to calculate $u_{i2}(\beta_1, \beta_2)$ and $w_{i2}(\beta_1, \beta_2)$. However, QCD solves the algorithm without replacing $u_{ij}(\beta_1, \beta_2)$ and does not change the objective function. \\

In this simulation, for simplicity, we use $n=5$ and estimate $\beta_1$ and $\beta_2$ sequentially by initializing both with $0$. In Figure \ref{fig:converge1} and \ref{fig:converge}, we report the objective function values and the estimated $\hat{\beta}_1, \hat{\beta}_2$ of the first two iterations for QCD and QICD. The first iteration is computing $L(\beta_1, 0)$ and using the estimated $\hat{\beta_1}$ to compute $L(\hat{\beta}_1, \beta_2)$. Second iteration is calculating $L(\beta_1, \hat{\beta}_2)$ then $L(\hat{\beta}_1, \beta_2)$. In this example, we could notice that the objective function that QICD and QCD are minimizing are different. In addition, we could say, for some cases, QICD estimates (blue dots) are different from the QCD estimates (red dots). By approximating the objective function, QICD is not minimizing the exact objective function, which might yield unexact CD solution in some scenarios.

\section{Exact Pathwise Coordinate Descent update for Non-convex Penalized Quantile Regression}\label{apdx:scad_mcp}

This part provides a step-by-step QCD algorithm for penalized quantile regression with non-convex penalties. The penalties considered are SCAD and MCP. Note that for these algorithms, we do not consider the KKT condition.

\subsection{MCP penalized quantile regression}\label{sec:mcp}

In this section of appendix, we provide preliminary results of exact CD update for MCP-penalized QR. In case of non-convex penalty MCP, the derivation logic of exact CD update for Lasso penalized QR maintains the same, but some variations are made. We first use the MCP penalty defined in \citet{peng2015iterative} and follow the identical steps from \eqref{obj_qrlasso} to \eqref{eqn:obj_qrlasso_simple}. Then, the MCP penalized QR problem for fixed $j$, $\lambda>0$, and $a>1$ is
\begin{align}\label{obj_mcp}
    \min_{\beta} &\sum_{i : x_{i}>0} |x_{i}|\rho_\tau(v_{i}-\beta)+ \sum_{i : x_{i}<0} |x_{i}|\rho_{1-\tau}(v_{i} - \beta)
     + \lambda \left( |\beta| - \frac{\beta^2}{2 a\lambda} \right) \mathbb{I}(0 \leq |\beta| < a \lambda) + \frac{a \lambda^2}{2} \mathbb{I}(|\beta| \geq a \lambda).
\end{align}

To track the derivatives of $\beta$ recursively within each range, we add $0$, $-a\lambda$, and $a\lambda$ to the $n$ $v_i$'s and order $n+3$ data points : $v_1<v_2<\ldots<v_k<-a\lambda<v_{k+2}<\ldots<v_l<0<v_{l+2}<\ldots<v_m<a\lambda<v_{m+2}<\ldots<v_{n+3}$. One thing to note here is that the $x_i$'s attached to the added $0, - a\lambda,$ and $a\lambda$ are zeros. \\

Starting from the first range $\beta<v_1$, we track the objective function value $S_0$ and its derivative $S_0'$. Since $v_1<-a\lambda$, we will get the penalty of ${a\lambda^2}/{2}$ within this range.
\begin{align*}
    \bullet \ &\beta < v_1 \\
    S_0 &= \sum_{i : x_i>0} |x_i| \tau (v_i-\beta) + \sum_{i : x_i<0} |x_i| (1-\tau) (v_i-\beta)  + \frac{a\lambda^2}{2} \\ &= \sum_{i=1}^{n} |x_i|  \left( \tau \mathbf{I}(x_i > 0) + (1-\tau)\mathbf{I}(x_i < 0)  \right) (v_i-\beta) + \frac{a\lambda^2}{2} \\ &= S + \frac{a\lambda^2}{2} \\
    S_0' &= -\sum_{i : x_i>0} |x_i| \tau -\sum_{i : x_i<0} |x_i| (1-\tau) \\ &= -\sum_{i=1}^{n} |x_i|  \left( \tau \mathbf{I}(x_i > 0) + (1-\tau)\mathbf{I}(x_i < 0)  \right) \\ &= S'
\end{align*}
For the range $v_1<\beta<v_2$, equivalent logic that $S_1$ and $S_1'$ are the same in either case $x_1>0$ or $x_1<0$ from Lasso penalized QR derivation maintains. The recursive formula with the penalty of $a\lambda^2/2$ carries on until $v_k<\beta<-a\lambda$.
\begin{align*}
    \bullet \ &v_1 < \beta < v_2 \\
     S_{1} &= |x_1| (\tau-1)(v_1-\beta) + S - |x_1|\tau(v_1-\beta)   + \frac{a\lambda^2}{2}  \\  &= S_0 - |x_1|(v_1-\beta) \\ 
     S_{1}' &= S_0' + |x_1| \\
     \smallskip \\
     \bullet \ &v_1 < \ldots < v_k <\beta < -a\lambda \\
     S_{k} &= S  - \sum_{i=1}^k |x_i| (v_i-\beta)  + \frac{a\lambda^2}{2}\\ &= S_{k-1} - |x_k|(v_k-\beta)  \\
     S_{k}' &= S_0' + \sum_{i=1}^k |x_i| \\ &= S_{k-1}' + |x_k|
\end{align*}

For $-a\lambda < \beta < v_{k+1}$, which is right after $\beta$ exceeds $-a\lambda$, we need to consider the penalty term when $0\leq|\beta|<a\lambda$. Note that $\beta<0$, so we will have $\lambda \left(-\beta - \frac{\beta^2}{2a\lambda}\right)$. When expressing $S_{k+1}$ in a recursive form, we need to subtract the previous penalty $a\lambda^2/2$ in order to substitute with the new one. This penalty persists until $v_1< \beta < 0$.
\begin{align*}
    \bullet \ v_1 &<\ldots <  -a\lambda < \beta < v_{k+2} \\
     S_{k+1} &= S - \sum_{i=1}^{k+1} |x_i| (v_i-\beta)  + \lambda \left( -\beta - \frac{\beta^2}{2a\lambda}\right) \\ &= S_{k} - |x_{k+1}|(v_{k+1}-\beta) -\frac{a\lambda^2}{2} + \lambda \left( -\beta - \frac{\beta^2}{2a\lambda}\right)  \\
     S_{k+1}' &= S' + \sum_{i=1}^{k+1} |x_i| + \left( -\lambda - \frac{\beta}{a}\right) \\ &= S_{k}' + |x_{k+1}| + \left( -\lambda - \frac{\beta}{a}\right) \\
     &\vdots \\
     \bullet \ v_1 &<\ldots < v_l < \beta < 0 \\
     S_{l} &= S_{l-1} - |x_{l}|(v_{l}-\beta)  \\
     S_{l}' &=  S_{l-1}' + |x_{l}|
\end{align*}

Once $\beta$ exceeds $0$, the penalty term becomes $\lambda \left(\beta - \frac{\beta^2}{2a\lambda}\right)$ from the range $0<\beta<v_{l+2}$ to $v_m<\beta<a\lambda$ since $|\beta|=\beta$. As we follow the logic of Lasso penalized QR, we add $2\lambda$ to the recursive form here also.

\begin{align*}
    \bullet \ v_1 &<\ldots < v_l < 0 < \beta < v_{l+2} \\
     S_{l+1} &= S - \sum_{i=1}^{l+1} |x_i| (v_i-\beta)  + \lambda \left(\beta - \frac{\beta^2}{2a\lambda}\right) \\ &= S_{l} - |x_{l+1}|(v_{l+1}-\beta)+ 2\lambda\beta  \\
     S_{l+1}' &= S' + \sum_{i=1}^{l+1} |x_i| + \left(\lambda - \frac{\beta}{a}\right) \\ &= S_{l}' + |x_{l+1}| + 2\lambda \\
     &\vdots \\
    \bullet \ v_1 &<\ldots < 0 < \ldots < v_m < \beta < a\lambda \\
     S_{m} &= S - \sum_{i=1}^{m} |x_i| (v_i-\beta)  + \lambda \left(\beta - \frac{\beta^2}{2a\lambda}\right) \\ &= S_{m-1} - |x_{m}|(v_{m}-\beta)  \\
     S_{m}' &= S' + \sum_{i=1}^{m} |x_i| + \left(\lambda - \frac{\beta}{a}\right) \\ &= S_{m-1}' + |x_{m}|
\end{align*}

We could observe that the derivative is not a constant but a line within the range $-a\lambda < \beta < a\lambda$ as in Figure \ref{fig:simpleQR_l1}. In addition, the slope of this line is always $-1/a$, which is negative. Suppose a derivative that is a line with negative slope crossing the x-axis. Then we could notice the x-intercept of this slope becomes a local maximum of the original convex function. Following this reasoning, the solution to the minimization problem \eqref{obj_mcp} will not occur at the x-intercept of the derivative line. Furthermore, the minimization will not take place within the range where the slope remains under the x-axis since the slope is decreasing. Therefore, we can obtain the solution through the identical rule as Proposition \ref{prop3}.\\

\begin{figure}[t!]
\begin{center}
\centerline{\includegraphics[width=\columnwidth]{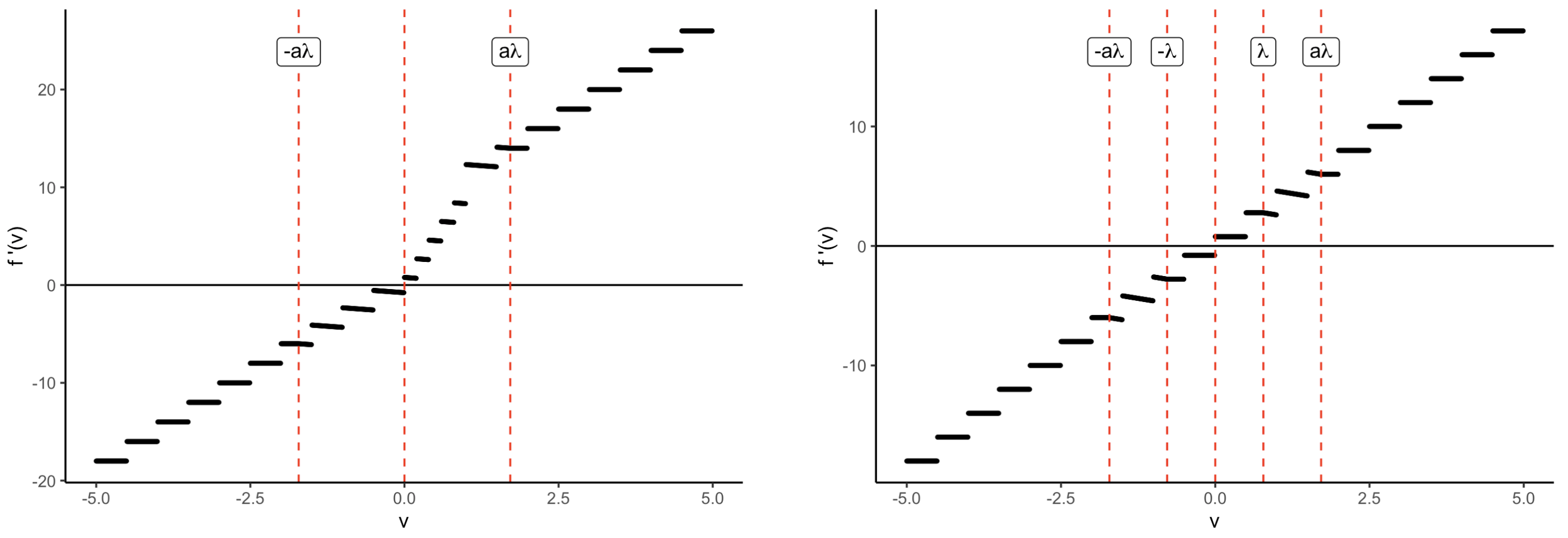}}
\caption{\textit{Derivatives of non-convex penalized QR.} (left) MCP penalized QR. (right) SCAD penalized QR. $\lambda=0.78$ and $a=2.2$ are used for both penalties.}
\label{fig:scad_mcp}
\end{center}
\vskip -0.2in
\end{figure}

In order to compute the derivatives within the range $-a\lambda<\beta<a\lambda$, we need to have a single derivative value. Suppose that the line derivatives are $f_i(\cdot)$. Then we can express the starting point of the line as a coordinate $\left( v_{i}, f_i(v_{i}) \right)$ on a plane. Hence, we will consider the derivative within each range between $-a\lambda$ and $a\lambda$ as $f_i(v_{i})$. By doing so, the recursive formula of $S_{k+1}$ and $S_{m}$ will contain extra calculation. These calculations will be offset once we replace $\beta$ with $-a\lambda$ and $a\lambda$. Keeping this in mind, we can reach the $\hat{\beta} = v_{t}$ by only tracing the derivatives.\\

After $\beta$ becomes larger than $a\lambda$, the objective function value and its derivative remains constant within each range and lasts until the end. Step-by-step algorithm is provided in Algorithm \ref{alg2}. Note that we do not include KKT condition check to preliminary results for the non-convex penalized QR problem. The shape of derivatives are depicted in Figure \ref{fig:scad_mcp}.

\begin{align*}
    \bullet \ v_1 &<\ldots < a\lambda < \beta < v_{m+1} \\
     S_{m} &= S - \sum_{i=1}^{m} |x_i| (v_i-\beta)  + \frac{a\lambda^2}{2} \\ &= S_{m-1} - |x_{m}|(v_{m}-\beta) - \lambda \left(\beta - \frac{\beta^2}{2a\lambda}\right) + \frac{a\lambda^2}{2}   \\
     S_{m}' &= S' + \sum_{i=1}^{m} |x_i| \\ &= S_{m-1}' + |x_{m}| - \left(\lambda - \frac{\beta}{a} \right) \\
     &\vdots \\
    \bullet \ v_1 &<\ldots < v_{n+3} < \beta  \\
     S_{n+3} &= S_{n+2} - |x_{n+3}|(v_{n+3}-\beta)  \\
     S_{n+3}' &=  S_{n+2}' + |x_{n+3}|
\end{align*}

\begin{algorithm}
\caption{Exact coordinate descent update for MCP penalized QR}
\label{alg2}
\begin{algorithmic}
    \STATE {\bfseries Input:} $X = \left(x_1 : x_2 : \ldots : x_n \right)^\top \in \mathbb{R}^{n \times p}$ \\ \smallskip \ \ \ \ \ \ \ \ \ \ \ \ \ $y = (y_1, \ldots, y_n) \in \mathbb{R}^n$ \\ \smallskip \ \ \ \ \ \ \ \ \ \ \ \ \ $\tau \in \left(0,1\right)$ \\ \smallskip \ \ \ \ \ \ \ \ \ \ \ \ \ $a > 1$ \\ 
    \smallskip \ \ \ \ \ \ \ \ \ \ \ \ \ $\lambda = \left(\lambda_1, \ldots, \lambda_L \right) \in \mathbb{R}^L$
   \STATE {\bfseries Result:} $\hat{\beta} = \left( \hat{\beta}^{(1)} : \hat{\beta}^{(2)} : \ldots : \hat{\beta}^{(L)} \right)^\top \in \mathbb{R}^{L \times p}$
   \STATE Initialize $\beta \gets \beta^{(0)} = \left( 0, \ldots, 0 \right)^\top$

   \FOR{$\ell=1$ {\bfseries to} $L$}
    \FOR{$j=1$ {\bfseries to} $p$}
        \STATE Compute and order \[v_{ij} = 
        \begin{dcases*}
        \frac{y_i-{\sum_{k \neq j} x_{ik}\hat{\beta}_k}}{x_{ij}}  & $i=1, \ldots, n$ \\ 0 & $i=n+1$ \\ -a\lambda_\ell & $i=n+2$ \\ a\lambda_\ell & $i=n+3$
        \end{dcases*}\]
        \STATE Define $S_{0,j}'=-\sum_{i=1}^n |x_{ij}| \left( \tau \mathbf{I}(x_{ij} > 0) + (1-\tau)\mathbf{I}(x_{ij} < 0) \right)$

    \STATE Compute cumulative sums $S_{0,j}'+|x_{1j}|, \ldots, S_{0,j}'+\sum_{i=1}^{n+3} |x_{ij}|$   
    \STATE Compute boost 
    \[\begin{dcases*}
    -\lambda_{\ell}-\frac{\beta_j}{a} &  $-a\lambda_\ell \leq \beta_j<0$ \\
    \lambda_\ell-\frac{\beta_j}{a} & $0 \leq \beta_j <a\lambda_\ell$ \\
    0 & $\lvert \beta_j \rvert \geq a\lambda_\ell$
    \end{dcases*}\]

    \STATE Add boost to each cumulative sum

    \STATE $\beta_j \gets \hat{\beta}_j=v_{t,j}$ \\
    $j \gets j+1$
   \ENDFOR


    \IF{nudge = TRUE}
    \STATE $\beta \gets \hat{\beta}^{(\ell)} + $ nudge
    \ENDIF

    \STATE $\beta \gets \hat{\beta}^{(\ell)}$
   \ENDFOR
\end{algorithmic}
\end{algorithm}

\subsection{SCAD penalized quantile regression}\label{sec:scad}

Another non-convex penalty that we consider is SCAD. The shape of SCAD is similar to MCP, but it contains Lasso penalty inside and gives larger constant penalty excessive $\beta$'s. The overall logic of obtaining derivatives is identical to Appendix \ref{sec:mcp}. Note again that we do not apply KKT condition check in non-convex penalized QR. By writing the SCAD penalized QR problem, we are able to recognize the difference from MCP penalized QR problem. For fixed $j$, $\lambda>0$ and $a>2$, we can write
\begin{align}\label{obj_scad}
    \min_{\beta} &\sum_{i : x_{i}>0} |x_{i}|\rho_\tau(v_{i}-\beta)+ \sum_{i : x_{i}<0} |x_{i}|\rho_{1-\tau}(v_{i} - \beta) \nonumber \\
     &+ \lambda |\beta| \mathbb{I}(0 \leq |\beta| < \lambda) + \frac{a\lambda|\beta| - (\beta^2 + \lambda^2)/2}{a-1} \mathbb{I}(\lambda \leq |\beta| \leq a \lambda) + \frac{(a+1) \lambda^2}{2} \mathbb{I}(|\beta| > a \lambda).
\end{align}
Based on our knowledge from the previous section, we can conjecture that we will have a constant derivative when $0\leq |\beta|<\lambda$ and $|\beta|>a \lambda$. Furthermore, looking at the power of $\beta$, we can guess that we will have a line as a derivative for $\lambda \leq|\beta|\leq a \lambda$. We will verify this by first including five additional data points $0, -a\lambda, -\lambda, \lambda,$ and $a\lambda$ and ordering : $v_1<\ldots<v_k<-a\lambda<v_{k+2}<\ldots<v_q<-\lambda<v_{q+2}<\ldots<v_l<0<v_{l+2}<\ldots<v_u<\lambda<v_{u+2}<\ldots<v_m<a\lambda<v_{m+2}<\ldots<v_{n+5}$. Once more, the $x_i$'s attached to these five points are zeros. Now, we illustrate the recursive formulas for derivative within each range. The penalty $(a+1)\lambda^2/2$ maintains until $\beta<-a\lambda$.
\begin{align*}
    \bullet \ &\beta < v_1 \\
    S_0 &= \sum_{i : x_i>0} |x_i| \tau (v_i-\beta) + \sum_{i : x_i<0} |x_i| (1-\tau) (v_i-\beta)  + \frac{(a+1)\lambda^2}{2} \\ &= \sum_{i=1}^{n} |x_i|  \left( \tau \mathbf{I}(x_i > 0) + (1-\tau)\mathbf{I}(x_i < 0)  \right) (v_i-\beta) + \frac{(a+1)\lambda^2}{2} \\ &= S + \frac{(a+1)\lambda^2}{2} \\
    S_0' &= -\sum_{i : x_i>0} |x_i| \tau -\sum_{i : x_i<0} |x_i| (1-\tau) \\ &= -\sum_{i=1}^{n} |x_i|  \left( \tau \mathbf{I}(x_i > 0) + (1-\tau)\mathbf{I}(x_i < 0)  \right) \\ &= S'\\
    \vdots \\
    \bullet \ &v_1 < \ldots < v_k <\beta < -a\lambda \\
     S_{k} &= S  - \sum_{i=1}^k |x_i| (v_i-\beta) + \frac{(a+1)\lambda^2}{2}  \\ &= S_{k-1} - |x_k|(v_k-\beta)  \\
     S_{k}' &= S_0' + \sum_{i=1}^k |x_i| \\ &= S_{k-1}' + |x_k|
\end{align*}
Once $-a\lambda < \beta < \lambda$, we replace the constant penalty to a second order penalty. 
As pointed out in \ref{sec:mcp}, we track $f_i(v_i)$ when we observe a line derivative. This will yield surplus calculation in the recursive form of the objective function when $\beta$ crosses specific thresholds $-a\lambda, -\lambda, 0, \lambda,$ and $a\lambda$. The surplus does not affect the objective function values and the derivatives once we plug in the threshold values to $\beta$. Moreover, the slope of the derivative in this range is fixed to a negative value $-1/(a-1)$, so we do not take the x-intercept of these lines into account when finding the solution of \eqref{obj_scad}.    
\begin{align*}
    \bullet \ v_1 &<\ldots <  -a\lambda < \beta < v_{k+2} \\
     S_{k+1} &= S - \sum_{i=1}^{k+1} |x_i| (v_i-\beta)  + \frac{-a\lambda\beta - (\beta^2 + \lambda^2)/2}{a-1} \\ &= S_{k} - |x_{k+1}|(v_{k+1}-\beta) -\frac{(a+1)\lambda^2}{2} - \frac{a\lambda\beta + (\beta^2 + \lambda^2)/2}{a-1}  \\
     S_{k+1}' &= S' + \sum_{i=1}^{k+1} |x_i| - \frac{a\lambda +\beta}{a-1} \\ &= S_{k}' + |x_{k+1}| -\frac{a\lambda +\beta}{a-1} \\
     &\vdots \\
     \bullet \ v_1 &<\ldots < -a\lambda < \ldots<v_q<\beta < -\lambda \\
     S_{q} &= S_{q-1} - |x_{q}|(v_{q}-\beta)  \\
     S_{q}' &=  S_{q-1}' + |x_{q}|
\end{align*}

Within the next range $-\lambda<\beta<\lambda$, the penalty becomes Lasso, so we can follow the same reasoning of Section \ref{sec:method}. The derivatives get $-\lambda$ boost when $-\lambda<\beta<0$ and $\lambda$ boost once $\beta$ crosses $0$. 
\begin{align*}
    \bullet \ v_1 &<\ldots < -\lambda < \beta < v_{q+2} \\
     S_{q+1} &= S - \sum_{i=1}^{q+1} |x_i| (v_i-\beta)  -\lambda\beta \\ &= S_{q} - |x_{q+1}|(v_{q+1}-\beta) + \frac{a\lambda\beta + (\beta^2 + \lambda^2)/2}{a-1}  -\lambda\beta \\
     S_{q+1}' &= S' + \sum_{i=1}^{q+1} |x_i| - \lambda \\ &= S_{q}' + |x_{q+1}| + \frac{a\lambda +\beta}{a-1} - \lambda \\
     &\vdots \\
     \bullet \ v_1 &<\ldots <v_l<\beta < 0 \\
     S_{l} &= S_{l-1} - |x_{l}|(v_{l}-\beta)  \\
     S_{l}' &=  S_{l-1}' + |x_{l}| \\
     \smallskip \\
     \bullet \ v_1 &<\ldots <0<\beta < v_{l+2} \\
     S_{l+1} &= S - \sum_{i=1}^{l+1} |x_i| (v_i-\beta)  + \lambda\beta \\
     &= S_{l} - |x_{l+1}|(v_{l+1}-\beta) + 2\lambda\beta \\
     S_{l+1}' &=  S_{l}' + |x_{l+1}| + 2\lambda \\
     &\vdots \\
     \bullet \ v_1 &<\ldots <v_u<\beta < \lambda \\
     S_{u} &= S_{u-1} - |x_{u}|(v_{u}-\beta)  \\
     S_{u}' &=  S_{u-1}' + |x_{u}|
\end{align*}

$\lambda<\beta<a\lambda$ suggests a line derivative with $|\beta|=\beta$ and the slope $-1/(a-1)$ does not change. 
\begin{align*}
    \bullet \ v_1 &<\ldots < \lambda < \beta < v_{u+2} \\
     S_{u+1} &= S - \sum_{i=1}^{u+1} |x_i| (v_i-\beta)  + \frac{a\lambda\beta - (\beta^2 + \lambda^2)/2}{a-1} \\ &= S_{u} - |x_{u+1}|(v_{u+1}-\beta) -\lambda\beta  + \frac{a\lambda\beta - (\beta^2 + \lambda^2)/2}{a-1}  \\
     S_{u+1}' &= S' + \sum_{i=1}^{u+1} |x_i| + \frac{a\lambda -\beta}{a-1} \\ &= S_{u}' + |x_{u+1}| - \lambda + \frac{a\lambda - \beta}{a-1}  \\
     &\vdots \\
     \bullet \ v_1 &<\ldots  < v_m < \beta < a\lambda \\
     S_{m} &= S_{m-1} - |x_{m}|(v_{m}-\beta)  \\
     S_{m}' &=  S_{m-1}' + |x_{m}|
\end{align*}

Beyond $a\lambda$, $\beta$ will get a constant penalty $(a+1)\lambda^2/2$ up to the end. 
\begin{align*}
    \bullet \ v_1 &<\ldots < a\lambda < \beta < v_{m+2} \\
     S_{m+1} &= S - \sum_{i=1}^{m+1} |x_i| (v_i-\beta)  + \frac{(a+1)\lambda^2}{2} \\ &= S_{m} - |x_{m+1}|(v_{m+1}-\beta) - \frac{a\lambda\beta - (\beta^2 + \lambda^2)/2}{a-1} + \frac{(a+1)\lambda^2}{2}  \\
     S_{m+1}' &= S' + \sum_{i=1}^{m+1} |x_i|  \\ &= S_{m}' + |x_{m+1}| - \frac{a\lambda - \beta}{a-1}  \\
     &\vdots \\
     \bullet \ v_1 &<\ldots  < v_{n+5} < \beta \\
     S_{n+5} &= S_{n+4} - |x_{n+5}|(v_{n+5}-\beta)  \\
     S_{n+5}' &=  S_{n+4}' + |x_{n+5}|
\end{align*}

\begin{algorithm}
\caption{Exact coordinate descent update for SCAD penalized QR}
\label{alg3}
\begin{algorithmic}
    \STATE {\bfseries Input:} $X = \left(x_1 : x_2 : \ldots : x_n \right)^\top \in \mathbb{R}^{n \times p}$ \\ \smallskip \ \ \ \ \ \ \ \ \ \ \ \ \ $y = (y_1, \ldots, y_n) \in \mathbb{R}^n$ \\ \smallskip \ \ \ \ \ \ \ \ \ \ \ \ \ $\tau \in \left(0,1\right)$ \\ \smallskip \ \ \ \ \ \ \ \ \ \ \ \ \ $a > 2$ \\ 
    \smallskip \ \ \ \ \ \ \ \ \ \ \ \ \ $\lambda = \left(\lambda_1, \ldots, \lambda_L \right) \in \mathbb{R}^L$
   \STATE {\bfseries Result:} $\hat{\beta} = \left( \hat{\beta}^{(1)} : \hat{\beta}^{(2)} : \ldots : \hat{\beta}^{(L)} \right)^\top \in \mathbb{R}^{L \times p}$
   \STATE Initialize $\beta \gets \beta^{(0)} = \left( 0, \ldots, 0 \right)^\top$

   \FOR{$\ell=1$ {\bfseries to} $L$}
    \FOR{$j=1$ {\bfseries to} $p$}
        \STATE Compute and order \[v_{ij} = 
        \begin{dcases*} 
    \frac{y_i-{\sum_{k \neq j} x_{ik}\beta_k}}{x_{ij}}  & $i=1, \ldots, n$ \\ 0 & $i=n+1$ \\ -a\lambda_\ell & $i=n+2$ \\ a\lambda_\ell & $i=n+3$ \\
    -\lambda_\ell & $i=n+4$ \\
    \lambda_\ell & $i=n+5$
    \end{dcases*}\]
        \STATE Define $S_{0,j}'=-\sum_{i=1}^n |x_{ij}| \left( \tau \mathbf{I}(x_{ij} > 0) + (1-\tau)\mathbf{I}(x_{ij} < 0) \right)$

    \STATE Compute cumulative sums $S_{0,j}'+|x_{1j}|, \ldots, S_{0,j}'+\sum_{i=1}^{n+5} |x_{ij}|$   
    \STATE Compute boost 
    \[\begin{dcases*}
    \frac{(a+1){\lambda_\ell}^2}{2} & $\lvert \beta_j \rvert > a\lambda_\ell$ \\
    \frac{-a{\lambda_\ell}-\beta_j}{a-1} & -$a\lambda_\ell \leq \beta_j \leq -\lambda_\ell$ \\
    -\lambda_\ell & $-\lambda_\ell<\beta_j \leq 0$ \\
    \lambda_\ell & $0 \leq \beta_j <\lambda_\ell$ \\
    \frac{a{\lambda_\ell}-\beta_j}{a-1} & $\lambda_\ell \leq \beta_j \leq a\lambda_\ell$  \end{dcases*}\]

    \STATE Add boost to each cumulative sum

    \STATE $\beta_j \gets \hat{\beta}_j=v_{t-1,j}$ \\
    $j \gets j+1$
   \ENDFOR


    \IF{nudge = TRUE}
    \STATE $\beta \gets \hat{\beta}^{(\ell)} + $ nudge
    \ENDIF

    \STATE $\beta \gets \hat{\beta}^{(\ell)}$
   \ENDFOR
\end{algorithmic}
\end{algorithm}

\section{Additional Visualizations}\label{apdx:figures}

In this part of the appendix, we record additional visualizations to supplement the results presented in Section \ref{sec:simulation}. \ref{apdx:regpath} expands the effect of nudge in \ref{subsec:regpath} by showing the regularization path of QCD and QICD in low dimension ($p=150, n=150$). We highlight the role of stopping rule \eqref{stoprule} in Section \ref{subsec:time} by reporting the average computation time plot and table when stopping rule is not used in \ref{apdx:comptime}. Lastly, we display the RMSE and AUROC plots mentioned in Section \ref{subsec:accuracy} for seven different dimensions $p=100, 300, 500, 700, 1000, 1500, 2000$ with fixed $n=300$ and expand on the results in \ref{apdx:acc}.

\subsection{Regularization path}\label{apdx:regpath}

\begin{figure}[t!]
\begin{center}
\centerline{\includegraphics[width=\columnwidth]{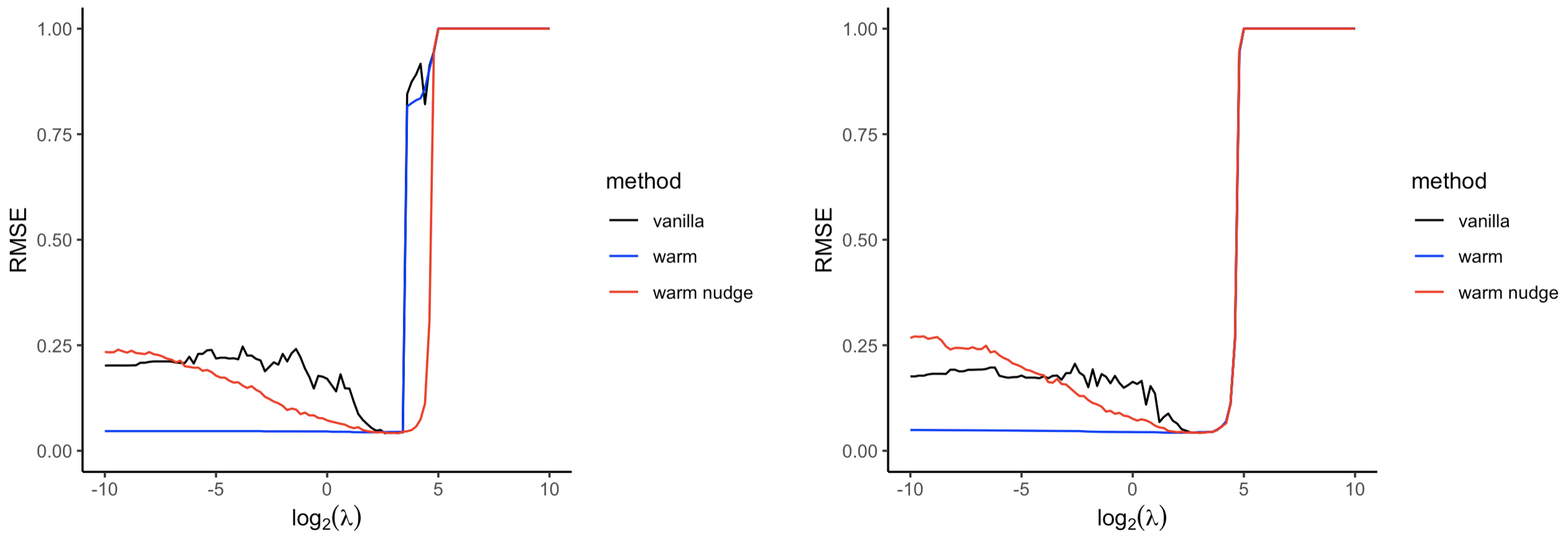}}
\caption{\textit{Regularization path of three methods per QCD and QICD algorithms in low dimension ($p = 150, n = 150$).} (left) RMSE plot of QCD. (right) RMSE plot of QICD.}
\label{fig:qcd_qicd_low}
\end{center}
\vskip -0.2in
\end{figure}

In Figure \ref{fig:qcd_qicd_low}, we report the regularization path of low dimension ($p=150, n=50$) data set for QCD and QICD algorithms. We observe similar patterns for vanilla version where the regularization path is bumpy for both algorithms as in Figure \ref{fig:qcd_qicd_high}. However, compared to Figure \ref{fig:qcd_qicd_high}, we notice that the regularization path for both warm QCD and warm QICD are stuck, but stuck in a stationary point where the minimum RMSE value of warm version is equivalent to the minimum RMSE of warm nudge version. 

\subsection{Computation time}\label{apdx:comptime}

\begin{figure}[ht]
\begin{center}
\centerline{\includegraphics[width=\columnwidth]{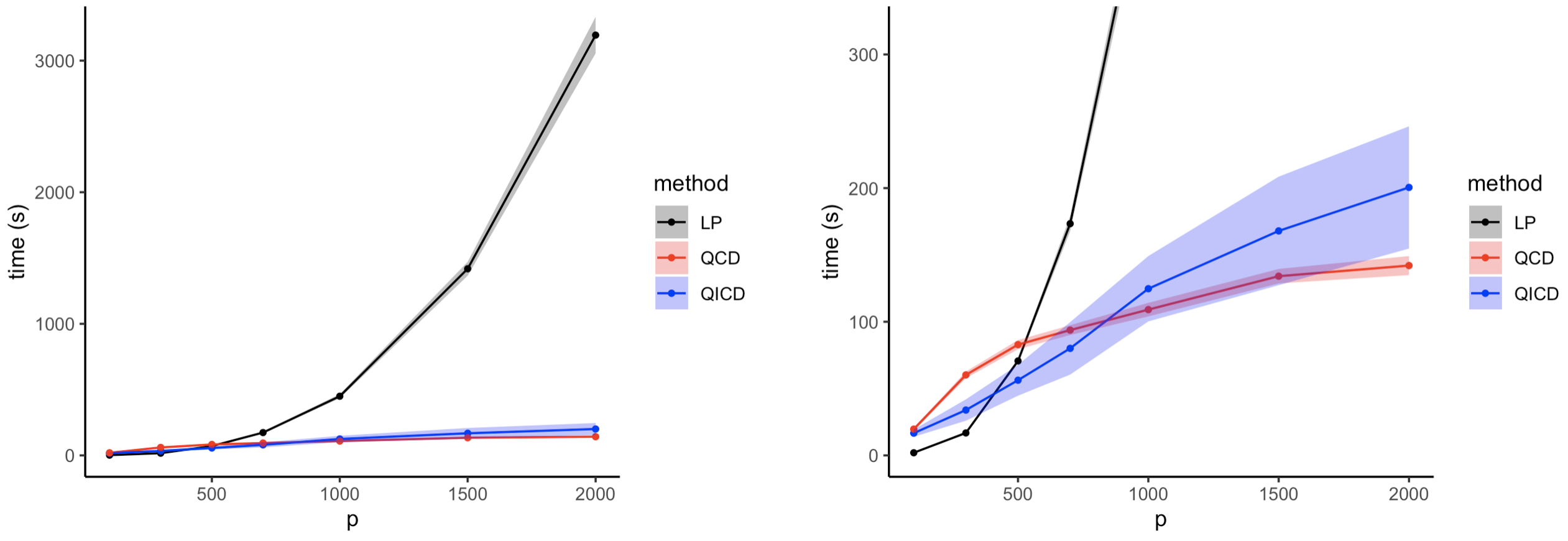}}
\caption{\textit{Average runtime of LP, QICD, and QCD without stopping rule.} (left) Full visualization of the average runtime. (right) Enlarged visualization of the average runtime up to 300 seconds.}
\label{fig:runtime_nostop}
\end{center}
\vskip -0.2in
\end{figure}

In Figure \ref{fig:runtime_nostop} and Table \ref{tab:runtime_nostop}, as dimension grows, we show that the computational time of LP grows linearly and it takes more than two times longer (3193.82 seconds) than when the stopping rule is used (1141.39 seconds) as in \ref{tab:runtime}. In addition, for low and moderately high dimensions $p=100, 300, 500, 700$, QICD seems to be faster than QCD. However, once we go to the ultra high dimension where $p=1000, 1500,$ and $2000$, QCD becomes much faster than QICD. \\

\begin{figure}[ht]
\begin{center}
\centerline{\includegraphics[width=\columnwidth]{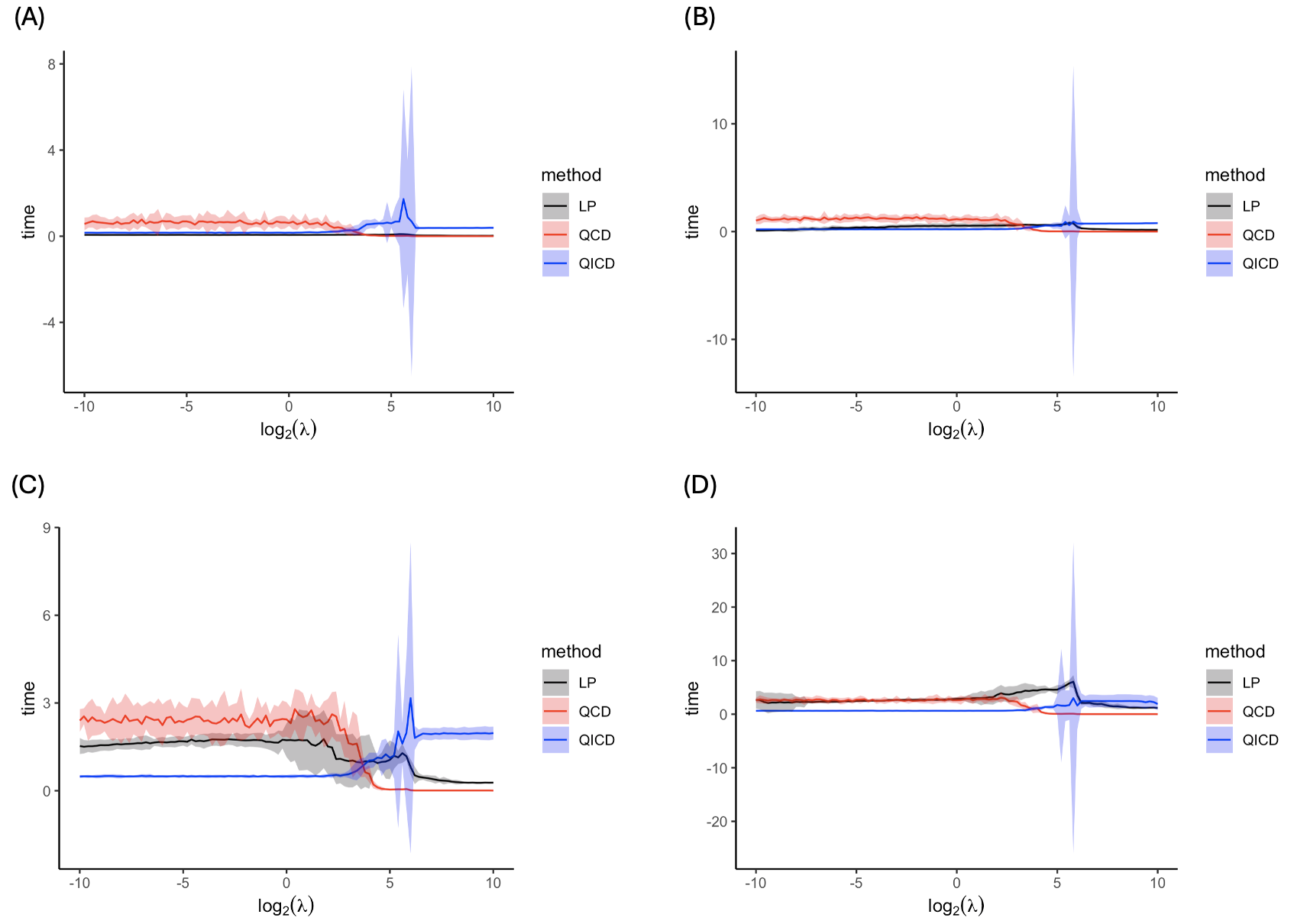}}
\caption{\textit{Median runtime at each $\lambda$ of LP, QICD, and QCD without stopping rule.} (A) $p=100$. (B) $p=300$. (C) $p=500$. (D) $p=700$. One interquartile range is computed over each dimension with fixed $n=300$. The results are the median of 20 different seeds.}
\label{fig:lambdatime}
\end{center}
\vskip -0.2in
\end{figure}

A thorough analysis of the runtime at low and moderately high dimensions are provided in Figure \ref{fig:lambdatime}. For all four plots in Figure \ref{fig:lambdatime}, QCD is the fastest when $\lambda$ is large which is when the solution is sparse. Once the solution becomes non-sparse, we see that QCD's computation time increases compared to QICD or LP. This also shows the merit of using the stopping rule. We acknowledge that giving a nudge is the reason for the above pattern. In the future, we plan to study how to adapt the value of nudge in a data-driven way.

\begin{table}[ht]
\tabcolsep=22pt
{\small
\resizebox{\textwidth}{!}{%
\begin{tabular}{c|ccc}
\hline
\multicolumn{1}{l}{} & \multicolumn{3}{|c}{time (seconds)}                \\
dimension (n = 300)           & LP              & QICD            & QCD             \\ \hline
p = 100                & 1.97 (0.04) & 16.64 (2.74)  & 19.64 (1.17) \\ 
p = 300                & 16.80 (0.97) & 33.94 (8.00) & 60.22 (2.51) \\
p = 500                & 70.66 (1.54) & 56.26 (11.59) & 82.89 (3.45) \\ 
p = 700                & 173.38 (5.79)  & 80.09 (19.64) & 93.76 (3.66) \\ 
p = 1000               & 449.70 (15.58) & 124.72 (24.46) & 109.08 (5.08) \\ 
p = 1500               & 1418.28 (52.88) & 167.96 (40.62) & 134.08 (5.42) \\ 
p = 2000               & 3193.82 (138.69) & 200.51 (45.72) & 142.05 (7.14) \\
\hline
\end{tabular}}%
}
\caption{\textit{Runtime, in seconds, of LP, QICD, and QCD for different dimensions $p$ when the stopping rule is not applied.} The standard deviations of runtime are reported inside the parentheses.}
\label{tab:runtime_nostop}
\end{table}

\subsection{Accuracy}\label{apdx:acc}

\begin{figure}[ht!]
\vskip 0.2in
\begin{center}
\centerline{\includegraphics[width=\columnwidth]{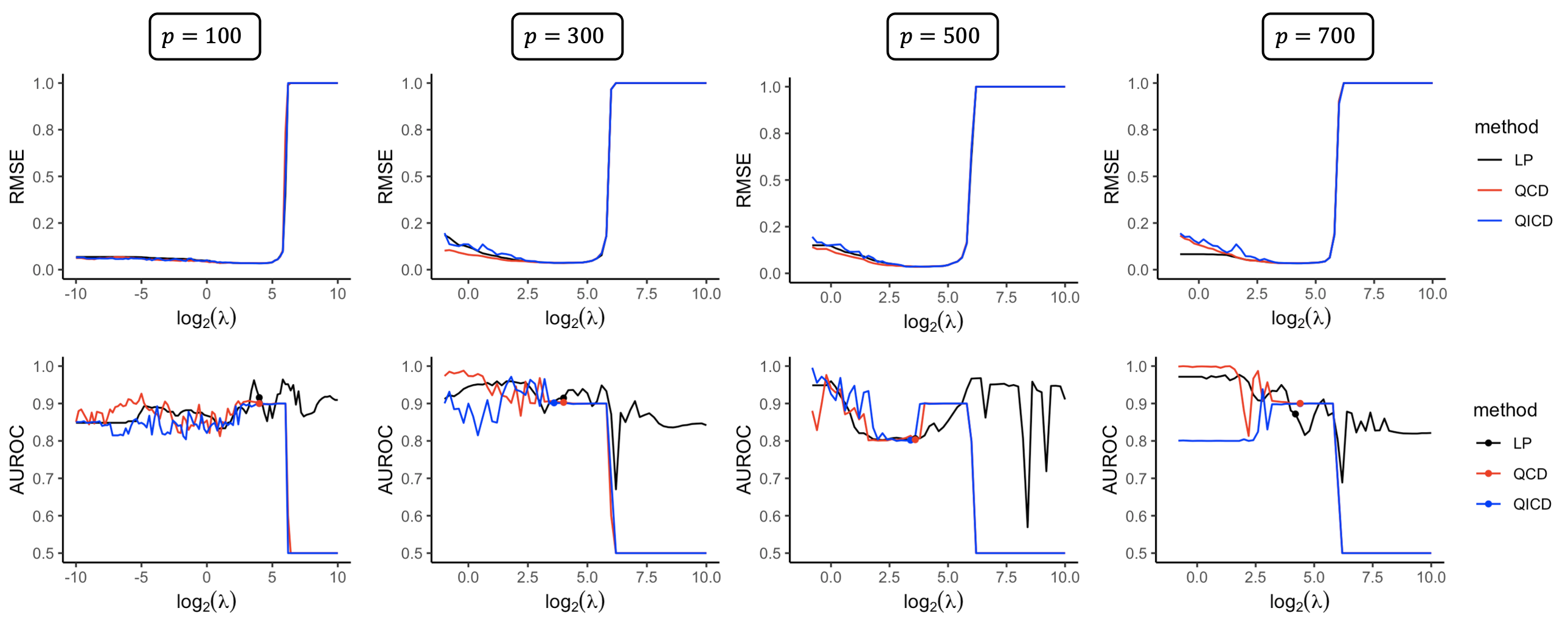}}
\centerline{\includegraphics[width=0.75\columnwidth]{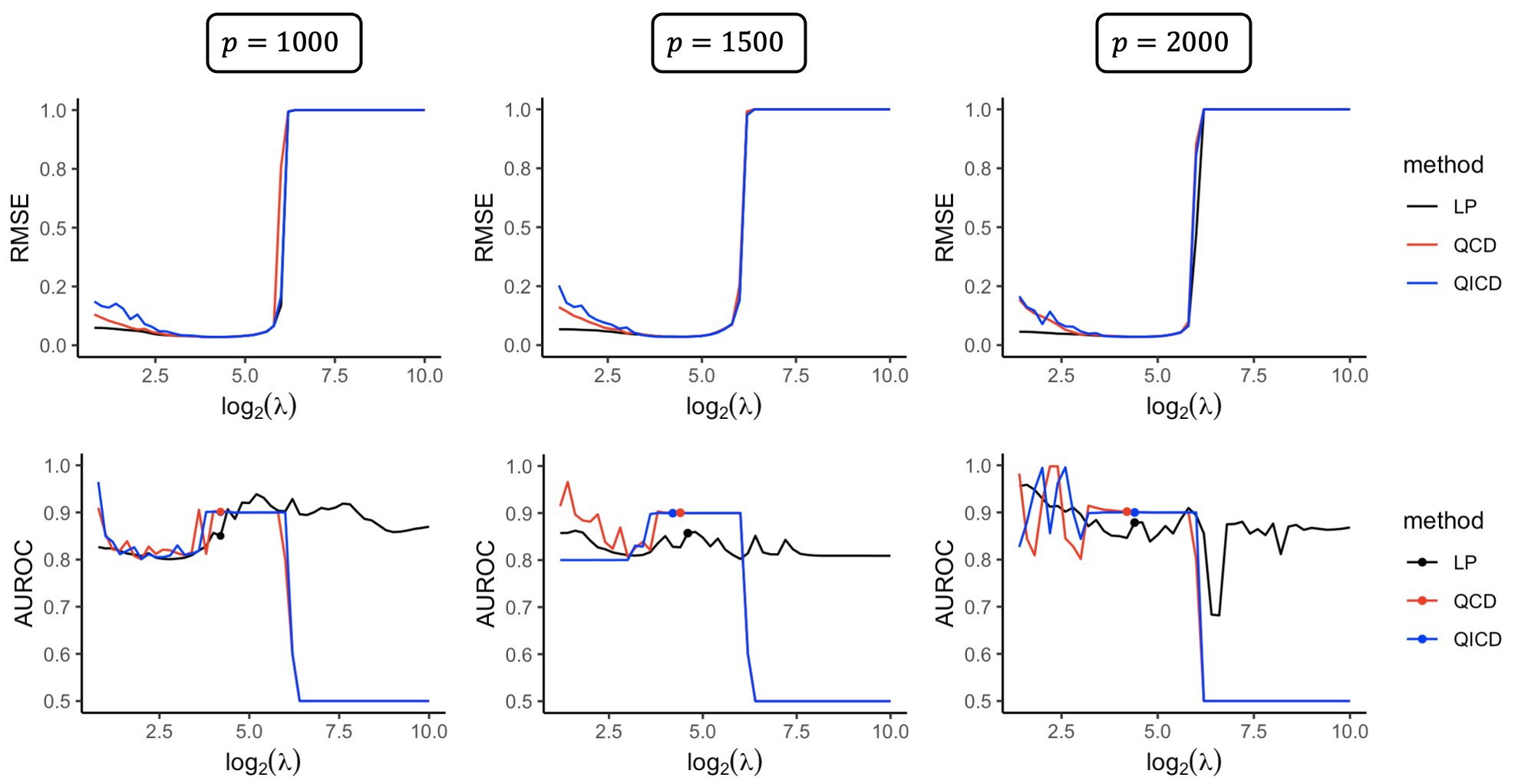}}
\caption{\textit{RMSE and AUROC plots of LP, QICD, and QCD for seven different dimensions}. Dots on the AUROC plot depicts the AUROC values of each method when minimum RMSE is achieved.}
\label{fig:rmse_auroc}
\end{center}
\vskip -0.2in
\end{figure}

We exhibit specific examples of the regularization paths and AUROC plots for $p=100, 300, 500, 700, 1000, 1500,$ and $2000$ in Figure \ref{fig:rmse_auroc}. The regularization paths for all three methods are stored on top and we see that they display a U shape. Note that we apply nudge to our QCD algorithm and the $\log_2(\lambda)$ grid is different due to the stopping rule for each dimension. The bottom shows the AUROC plots where the dots display the AUROC value when minimum RMSE is achieved for each method. Every AUROC plots confirm that QCD achieves higher or similar AUROC values as LP and QICD, indicating the estimation and model selection accuracy of QCD is competitive to existing methods.

